\documentclass[11pt]{article}
\usepackage{amsmath}
\usepackage{amssymb}
\usepackage{amsthm,amsxtra}
\usepackage{epsf}
\usepackage{eepic}
\newlength{\mytopmargin}
\newlength{\myleftmargin}
\newtheorem{lemma}{Lemma}
\newtheorem{prop}{Proposition}
\newtheorem{theorem}{Theorem}
\newtheorem{conj}{Conjecture}
\newcommand{\zz}{\mathbb Z}

\begin{document}
\vspace{2cm}
\noindent
\begin{center}{\Large \bf Applications and generalizations of
Fisher-Hartwig asymptotics} 
\end{center}
\vspace{5mm}

\begin{center}
P.J.~Forrester${}^*$ and N.E.~Frankel${}^\dagger$
\end{center}

\vspace{.2cm}

\noindent
${}^*$Department of Mathematics and Statistics, University of
Melbourne, Victoria 3010, Australia; \\
${}^\dagger$School of Physics, University of
Melbourne, Victoria 3010, Australia; \\
Electronic addresses: p.forrester@ms.unimelb.edu.au;
n.frankel@physics.unimelb.edu.au \\

\vspace{.2cm}
\small
\begin{quote}
Fisher-Hartwig asymptotics refers to the large $n$ form of a
class of Toeplitz determinants with singular generating functions.
This class of Toeplitz determinants occurs in the study of the 
spin-spin correlations for the two-dimensional Ising model, and the
ground state density matrix of the impenetrable Bose gas, amongst
other problems in mathematical physics. We give a new application
of the original Fisher-Hartwig formula to the asymptotic decay of the
Ising correlations above $T_c$, while the study of the Bose gas density
matrix leads us to generalize the Fisher-Hartwig formula to the asymptotic 
form of random matrix averages over the classical groups and the Gaussian
and Laguerre unitary matrix ensembles. Another viewpoint of our
generalizations is that they extend to Hankel determinants the
Fisher-Hartwig asymptotic form known for Toeplitz determinants.
\end{quote}

\section*{Encomium}
In celebration of Freeman Dyson on his
$$
\Big ( 32 ( {\scriptstyle {40960001 \over 25600}} )^3 - 6
({\scriptstyle {40960001 \over 25600}}) +
\sqrt{[32  ( {\scriptstyle {40960001 \over 25600}} )^3 - 6
({\scriptstyle {40960001 \over 25600}} )]^2 - 1 } \Big )^{1/6} \: {\rm th}
$$
birthday\footnote{This radical appears in the works on Ramanujan
\cite{BCZ97}, who is very dear to Dyson.}.

Dyson's legendary works \cite{Dy62} on random matrices are now
standards in physics, mathematics and fields far(ther) afield. These
works and, in particular his powerful log-Coulomb gas model, developed
to liberate the mathematics where none yet exists, are the very
essential tools in our ongoing pursuits \cite{Fo02}.

\section{Introduction}
\setcounter{equation}{0}
Fisher-Hartwig asymptotics refers to the large $n$ form of a class of
Toeplitz determinants $D_n[g]$. By definition, the entries of the latter
depend only on the difference of the row and column indices, and thus
\begin{equation}\label{1}
D_n[g] = \det [g_{j-k}]_{j,k=1,\dots,n}
\end{equation}
for some $\{g_{k}\}_{k=0,\pm1,\pm2,\dots}$. Crucial to the structure of the
asymptotic form of (\ref{1}) are analytic properties of the so called
symbol
\begin{equation}\label{1.1}
g(\theta) := \sum_{n=-\infty}^\infty g_n e^{i n \theta},
\end{equation}
or more particularly the decay of the Fourier coefficients of
$\log g(\theta)$. Explicitly, let
\begin{equation}\label{2}
\log g(\theta) = \sum_{p=-\infty}^\infty c_p e^{ip \theta}.
\end{equation}
Then if
\begin{equation}\label{3}
\sum_{p=-\infty}^\infty |p| c_p c_{-p} < \infty
\end{equation}
a strong form of the Szeg\"o limit theorem (see e.g.~\cite{Wi76,Jo88})
asserts that for $n \to \infty$
\begin{equation}\label{4}
D_n[g] = \exp \Big ( n c_0 + \sum_{k=1}^\infty k c_k c_{-k} +
{\rm o}(1) \Big ).
\end{equation}

Two cases for which (\ref{3}) will not hold are when $g(\theta)$ has a 
jump discontinuity or a zero for some $-\pi < \theta \le \pi$. It is for
such singular symbols (in the case of a zero it is the logarithm of 
the symbol which is singular) that Fisher and Hartwig \cite{FH68}
sought the asymptotic form of (\ref{1}). Symbols with singularities of
this type have the functional form
\begin{eqnarray}\label{5}
\log g(\theta) & = & \log a(\theta) - i \sum_{r=1}^R b_r
{\rm arg} \, e^{i (\theta_r + \pi - \theta)} +
 \sum_{r=1}^R a_r \log | 2 - 2 \cos (\theta - \theta_r) | \nonumber \\
 & = & \log a(\theta) + \sum_{r=1}^R \Big ( (a_r + b_r)
\log (1 + e^{i (\theta - (\theta_r + \pi))}) +
(a_r - b_r) \log (1 + e^{i (\theta_r + \pi - \theta)})   \Big ).
\end{eqnarray}
Here $- \pi < {\rm arg} \, z \le \pi$ and $a(\theta)$ is assumed to be
sufficiently smooth that if we write
\begin{equation}\label{1.6a}
\log a(\theta) = \sum_{p=-\infty}^\infty c_p e^{i p \theta}
\end{equation}
(cf.~(\ref{2})) then the condition (\ref{3}) holds. By using data following
from the fact that special cases of (\ref{5}) correspond to Toeplitz
determinant expressions for the spin-spin correlation in the
two-dimensional Ising model at criticality (see Section 2 below), the
asymptotic form of which had previously been calculated \cite{Wu66},
Fisher and Hartwig \cite{FH68} conjectured that for some range of
parameter values $\{a_r\}_{r=1,\dots,R}$,
$\{b_r\}_{r=1,\dots,R}$,
\begin{equation}\label{FH}
D_n[g] \mathop{\sim}\limits_{n \to \infty} e^{c_0 n}
e^{\sum_{r=1}^R (a_r^2 - b_r^2) \log n} E
\end{equation}
where $E$ is independent of $n$. Subsequently this was proved for various
ranges of parameter values (see e.g.~\cite{BK94}) and furthermore the constant
was determined to be given by
\begin{eqnarray}\label{FH1}
E & = & e^{\sum_{k=1}^\infty k c_k c_{-k} }
\prod_{r=1}^R e^{-(a_r + b_r) \log a_-(\theta_r) }
 e^{-(a_r - b_r) \log a_+(\theta_r) } \nonumber \\
 &  & \times \prod_{1 \le r \ne s \le R}
(1 - e^{i (\theta_s - \theta_r)})^{-(a_r+b_r)(a_s-b_s)}
\prod_{r=1}^R {G(1+a_r +b_r) G(1+a_r - b_r) \over
G(1+2 a_r) }
\end{eqnarray}
where $G$ is the Barnes $G$-function and
\begin{equation}\label{8.1}
 \log a_+(\theta) := \sum_{p=1}^\infty c_p e^{i p \theta}, \qquad
\log a_-(\theta) := \sum_{p=-\infty}^{-1} c_p e^{i p \theta}.
\end{equation}

Our interest is in applications and generalizations of the Fisher-Hartwig
asymptotic formula (\ref{FH}). We begin in Section 2 with an application
of (\ref{FH}) to the calculation of the asymptotic form of the 
spin-spin correlation for the two-dimensional Ising model above
criticality. In Section 3 the well known equivalence of the 
Toeplitz determinant (\ref{1}) to a random matrix average over the
unitary group $U(n)$ is revised. This average is in turn equivalent to the
partition function of the one-component log-gas on a circle, subject to
a one-body potential with Boltzmann factor $g(\theta)$ at the special
coupling $\beta = 2$. As such there is a natural generalization for 
couplings $\beta > 0$, and in the case $b_r = 0$, $r=1,\dots,R$ this can be
used to predict the corresponding generalization of (\ref{FH}).
Moreover, in the special case $a(\theta) = 1$, $R=1$ the sought asymptotic
form can be deduced from an exact formula valid for general $a_r, b_r$. 
This can be used to extend the conjectured generalization of (\ref{FH})
to non-zero $b_r$.

In Section 4 we recall the problem of computing the asymptotic
form of the density matrix for impenetrable bosons in Dirichlet
and Neumann boundary conditions. This is immediately identifiable as an
average over the classical groups $Sp(N)$ and $O^+(2N)$ respectively, with
the function being averaged over having two zeros, and thus analogous to the
random matrix formulation of the Toeplitz determinant (\ref{1}) with
symbol (\ref{5}) in the case $R=2$, $b_r=0$. 
We point out that the same class of averages over the groups $O^+(2N+1)$
or $O^-(2N+1)$ result from considering the density matrix for the
impenetrable Bose gas in the case of mixed Dirichlet and 
Neumann boundary conditions. 
In \cite{FFG03} the sought
asymptotics were calculated on the basis of a combination of analytic
and log-gas arguments, and a Fisher-Hartwig type generalization
(with $b_r=0$) conjectured. The conjecture of
\cite{FFG03} can used to predict the asymptotic form in the case
of mixed Dirichlet and Neumann boundary conditions. Moreover we show
that this asymptotic
form can be proved
by making use of
asymptotic formulas recently obtained \cite{BE02} for
Toeplitz $+$ Hankel determinants
\begin{equation}\label{1.9a}
\det [ a_{j-k} + a_{j+k+1}  ]_{j,k=0,\dots,n-1}
\end{equation}
in the case of singular generating functions (\ref{5}). 

In addition to averages over the classical groups, the study of 
the density matrix for impenetrable bosons naturally leads
to the question of obtaining the asymptotic form of averages over the
eigenvalue probability density function for the GUE and LUE, in the case
that the function being averaged over has zeros. Here the GUE denotes
the Gaussian unitary ensemble of random Hermitian matrices, and the
LUE denotes the Laguerre unitary ensemble of positive definite matrices
with complex entries. 
These random matrix averages are equivalent to pure Hankel determinants
\begin{equation}\label{1.9b}
\det [a_{j+k}]_{j,k=0,\dots,n-1}, \qquad
a_n = \int_{-\infty}^\infty a(x) x^n \, d\mu(x)
\end{equation}
where $d\mu(x) = e^{-x^2} dx$ for the GUE and $d\mu(x) = x^a e^{-x} dx$,
$x>0$ for the LUE.
Conjectures for such asymptotic forms are given in
Section 5. The paper ends with some concluding remarks on the universal
form for Hankel asymptotics in Section 6, and attention is also drawn to
the fluctuation formula perspective of our asymptotic results.

\section{Spin-spin correlations for the two-dimensional Ising model}
\setcounter{equation}{0}
In the two-dimensional Ising model on a square lattice each site $(i,j)$ of 
the lattice exists in one of two possible states $\sigma_{ij} = \pm 1$
with coupling between nearest neighbours in the horizontal and vertical
directions. Explicitly, the joint probability density function for a 
particular configuration $\{\sigma_{ij} \}$ of the states on a
$(2N+1) \times (2N+1)$ lattice is given by
\begin{equation}\label{O1}
P_{2N+1}(\{\sigma_{ij}\}) = {1 \over Z_{2N+1} }
\exp \Big ( K_1 \sum_{j=-N}^{N} \sum_{i=-N}^{N-1} \sigma_{ij} \sigma_{i+1 \,j} +
K_2 \sum_{i=-N}^N \sum_{j=-N}^{N-1} \sigma_{ij} \sigma_{i \,j+1} \Big )
\end{equation}
where $Z_{2N+1}$ is the normalization. The spin-spin correlation function
between the spin $\sigma_{00}$ at the centre of the lattice, and the spin
$\sigma_{i^* j^*}$ at site $(i^*,j^*)$ is, in the infinite lattice limit,
defined as
\begin{equation}\label{O2}
\langle \sigma_{00} \sigma_{i^* j^*} \rangle = \lim_{N \to \infty}
\sum_{\{\sigma_{ij}\}} \sigma_{00} \sigma_{i^* j^*} P_{2N+1}(\{\sigma_{ij}\}).
\end{equation}

Onsager knew of, but never published (see instead e.g.~\cite{MW73}) a
Toeplitz determinant form for the case of (\ref{O2}) for which
$(i^*,j^*) = (n,n)$ and thus lies on the diagonal. Explicitly
\begin{equation}\label{h1}
\langle \sigma_{00} \sigma_{nn} \rangle = \det [a_{i-j} ]_{i,j=1,\dots,n},
\qquad a_p = {1 \over 2 \pi} \int_{-\pi}^\pi h(\theta) e^{-ip \theta} \,
d \theta
\end{equation}
where
\begin{equation}\label{h2}
h(\theta) := \Big ( {1 + (1/k) e^{-i \theta} \over
1 + (1/k) e^{i \theta} } \Big )^{1/2}, \qquad k = \sinh 2K_1 \sinh 2K_2.
\end{equation}
Also, in the case of (\ref{O2}) with $(i^*,j^*) = (0,n)$ so that the two
spins lie in the same row, Onsager and Kaufmann \cite{KO49} expressed
(\ref{O2}) as the sum of two Toeplitz determinants. A different approach
to this problem was undertaken by Potts and Ward \cite{PW55}, who obtained
instead the single Toeplitz determinant form
\begin{equation}\label{h3}
\langle \sigma_{00} \sigma_{0n} \rangle = \det [\tilde{a}_{i-j} 
]_{i,j=1,\dots,n},
\qquad \tilde{a}_p = {1 \over 2 \pi} \int_{-\pi}^\pi 
\tilde{h}(\theta) e^{-ip \theta} \,
d \theta
\end{equation}
where
\begin{equation}\label{h4}
\tilde{h}(\theta) := \Big ( {(1 + \alpha_1 e^{i \theta} )
(1 + \alpha_2 e^{-i \theta}) \over (1 + \alpha_1 e^{-i \theta} )
(1 + \alpha_2 e^{i \theta}) } \Big )^{1/2}
\end{equation}
with
$$
\alpha_1 := e^{-2 K_2} \tanh K_1, \qquad \alpha_2 := {e^{-2 K_2} \over
\tanh K_1 }.
$$
The formula obtain in  \cite{KO49} was shown to be identical to (\ref{h1}),
(\ref{h2})
by Montroll, Potts and Ward \cite{MPW63}. We remark that if $\alpha_1,
\alpha_2$ in (\ref{h4}) and $k$ in (\ref{h2}) are regarded as parameters
not specified by $K_1, K_2$, then setting $\alpha_1=0, \alpha_2 = 1/k$
in the former gives (\ref{h2}).

A detailed study of the asymptotic form of (\ref{h4}) was undertaken by
Wu \cite{Wu66}. Indeed, it was the asymptotic form of (\ref{h3}) at the
critical coupling
\begin{equation}\label{a12}
\alpha_1 < \alpha_2 = 1
\end{equation}
obtained in \cite{Wu66} which, partially at least, inspired the
formulation of the Fisher-Hartwig asymptotic formula (\ref{FH})
\cite{FH68}. To see how (\ref{FH}) relates to (\ref{h3}) with parameters
(\ref{a12}), note
\begin{equation}\label{a13}
\log \tilde{h}(\theta) \Big |_{\alpha_2 = 1} =
\log \Big ( {1 + \alpha_1 e^{i \theta} \over 1 + \alpha_1 e^{-i \theta} }
\Big )^{1/2} + i {\rm arg} \, e^{-i \theta / 2}.
\end{equation}
For $|\alpha_1| < 1$ this has the structure of (\ref{5}) with
$$
a(\theta) = \Big ( {1 + \alpha_1 e^{i \theta} \over 1 + \alpha_1 e^{-i \theta} }
\Big )^{1/2}, \: \: R=1, \: \: b_r = - {1 \over 2}, \: \: a_r=0, \: \:
\theta_r = - \pi.
$$
Recalling the definitions (\ref{2}) (with $g(\theta)$ replaced by $a(\theta)$)
and (\ref{8.1}), application of (\ref{FH}) implies
\begin{equation}\label{a14}
\langle \sigma_{00} \sigma_{0n} \rangle \Big |_{\alpha_2 = 1 \atop
\alpha_1 < 1} \mathop{\sim}\limits_{n \to \infty} 
\Big ( {1 + \alpha_1 \over 1 - \alpha_1} \Big )^{1/4}
{\sqrt{\pi} G^2(1/2) \over n^{1/4} }
\end{equation}
where use has been made of the functional equation
$$
G(z+1) = \Gamma(z) G(z),
$$
in agreement with the result of Wu.

The high temperature phase corresponds to couplings
\begin{equation}\label{b12}
\alpha_1 < 1 < \alpha_2, \qquad \alpha_1 \alpha_2 < 1.
\end{equation}
In this case $\log \tilde{h}(\theta)$ is of the form (\ref{5}) with
\begin{equation}\label{h5}
a(\theta) =
 \Big ( {(1 + \alpha_1 e^{i \theta} )
(1 + e^{i \theta}/\alpha_2)  \over (1 + \alpha_1 e^{-i \theta} )
(1 + e^{-i \theta}/\alpha_2 ) } \Big )^{1/2}, \: \:
R=1, \: \: b_r=-1, \: \: a_r=0, \: \: \theta_r = - \pi.
\end{equation}
With $R=1, b_r=-1, a_r=0$ we see that the Fisher-Hartwig asymptotic formula
(\ref{FH}) breaks down because according to (\ref{FH1}) the constant $E$
contains the factor $G(0) = 0$ and thus vanishes. To obtain the asymptotics
in this case the approach taken in \cite{Wu66} was to relate it back to the
original strong Szeg\"o theorem, multiplied by an auxilarly factor. Here we
will show that by transforming (\ref{h4}), a form of $\log \tilde{h}(\theta)$
can be obtained which has the general structure (\ref{5}) but is distinct 
from the specification (\ref{h5}). We will see that
applying the Fisher-Hartwig formula
then correctly reproduces the result of Wu for the leading asymptotic decay
in the high temperature phase.

For this purpose, let us introduce the notation $f(\theta) \equiv g(\theta)$
to mean that
$$
\int_{-\pi}^\pi f(\theta) e^{-i p \theta} \, d \theta =
c^{-p}\int_{-\pi}^\pi g(\theta) e^{-i p \theta} \, d \theta
$$
for some $c$ independent of $p$. According to the definitions
(\ref{h4}) and (\ref{h5}) we have $\tilde{h}(\theta) 
= e^{-i \theta} a(\theta)$. Now, since with $z = e^{i \theta} $,
$\tilde{h}(\theta)$ is an analytic function of $z$
in the annulus $1/\alpha_2 < |z| < \alpha_2$, by
Cauchy's theorem
$$
{1 \over 2 \pi}
\int_{-\pi}^\pi \tilde{h}(\theta) e^{-i p \theta} \, d \theta =
\int_{\cal C} \tilde{h}(\theta) z^{-p} {dz \over 2 \pi i z}
$$
for any simple closed contour encircling the origin in this annulus.
Choosing ${\cal C}$ to be the circle with radius $\alpha_2$ (the outer
boundary of the annulus) shows 
\begin{eqnarray*}
\tilde{h}(\theta) & \equiv & {e^{- i \theta} \over \alpha_2}
\Big ( {1 + \alpha_1 \alpha_2 e^{i \theta} \over 1 + ( \alpha_1 / \alpha_2)
e^{- i \theta} } \Big )^{1/2}
{ (1 + e^{i \theta} )^{1/2} \over (1 + e^{- i \theta}/ \alpha_2^2)^{1/2} }
\\ & = & {1 \over \alpha_2} {1 \over  (1 + e^{- i \theta}/ \alpha_2^2)^{1/2} }
\Big ( {1 + \alpha_1 \alpha_2 e^{i \theta} \over 1 + ( \alpha_1 / \alpha_2)
e^{- i \theta} } \Big )^{1/2} e^{-3 i \theta/4} |1 + e^{i \theta} |^{1/2}.
\end{eqnarray*}
This is of the form (\ref{5}) with
\begin{equation}\label{11'}
a(\theta) = {1 \over \alpha_2} {1 \over  (1 + e^{- i \theta}/ \alpha_2^2)^{1/2} }
\Big ( {1 + \alpha_1 \alpha_2 e^{i \theta} \over 1 + ( \alpha_1 / \alpha_2)
e^{- i \theta} } \Big )^{1/2}, \: \: R=1, \: \: b_r = -{3 \over 4}, \: \:
a_r = {1 \over 4}, \: \: \theta_r = - \pi.
\end{equation}
Application of (\ref{FH}) implies
\begin{equation}\label{12}
\langle \sigma_{00} \sigma_{0n} \rangle \Big |_{\alpha_1 <  1 < \alpha_2 \atop
\alpha_1 \alpha_2 < 1} \mathop{\sim}\limits_{n \to \infty}
{\alpha_2^{-n} \over (\pi n)^{1/2} } (1 - \alpha_1^2)^{1/4}
(1 - \alpha_2^{-2})^{-1/4} (1 - \alpha_1 \alpha_2)^{-1/2}
\end{equation}
in agreement with the result of Wu \cite{Wu66}. Moreover the Fisher-Hartwig
formula (\ref{FH}) with $R=1$ has been proved \cite{BS86} for parameter values 
satisfying all three of the inequalities
$$
{\rm Re} \, a_1 \ge 0, \qquad
{\rm Re} \, a_1 + {\rm Re} \, b_1 > -1, \qquad
{\rm Re} \, a_1 - {\rm Re} \, b_1 > -1.
$$
These inequalities are satisfied by the parameters in (\ref{11'}) and so the
Fisher-Hartwig formula provides a proof of (\ref{12}).

\section{$\beta$-generalization of the Fisher-Hartwig formula}
\setcounter{equation}{0}
It is well known, and easy to verify, that the Toeplitz determinant
(\ref{1.1}) can be written as a random matrix average according to
\begin{equation}\label{e.1}
D_n[g] = \Big \langle \prod_{l=1}^n g(\theta_l) \Big \rangle_{U(n)}.
\end{equation}
Here $U(n)$ refers to the eigenvalue probability density function for
the unitary group
\begin{equation}\label{e.2}
{1 \over (2 \pi)^n n!} \prod_{1 \le j < k \le n}
| e^{i \theta_k} - e^{i \theta_j} |^2, \qquad
- \pi < \theta_l \le \pi.
\end{equation}
As first noted by Dyson \cite{Dy62}, (\ref{e.2}) is proportional to the
Boltzmann factor for the one-component log-potential Coulomb gas on a 
circle, at the special coupling $\beta = 2$. From the log-gas viewpoint a
natural generalization of (\ref{e.2}) is the probability density function
C$\beta$E${}_n$ proportional to the Boltzmann factor for the same
statistical mechanical system but with general coupling $\beta > 0$,
\begin{equation}\label{e.3}
{1 \over (2 \pi)^n C_{n,\beta} } \prod_{1 \le j < k \le n}
| e^{i \theta_k} - e^{i \theta_j} |^\beta, \qquad
 C_{n,\beta} = {\Gamma(n\beta/2 + 1 ) \over (\Gamma(\beta/2+1))^n }.
\end{equation}
The identity (\ref{e.1}) then allows us to formulate a $\beta$-generalization
of the Toeplitz determinant (\ref{1.1}) as the average
\begin{equation}\label{e.4}
D_n^{(\beta)}[g] :=  \Big 
\langle \prod_{l=1}^n g(\theta_l) \Big \rangle_{{\rm C}\beta
{\rm E}_n}.
\end{equation}
Choosing $g(\theta)$ according to (\ref{5}) in this we obtain a natural
$\beta$-generalization of the Toeplitz determinant with a Fisher-Hartwig
symbol. In the case $b_r=0$, $r=1,\dots,R$, the log-gas viewpoint can
be used to conjecture the corresponding analogue of the asymptotic
formula (\ref{FH}).

Let
\begin{equation}\label{3.4'}
Z_n^{(\beta)}[g(\theta)] := {1 \over (2 \pi)^n}
\int_{-\pi}^\pi d \theta_1 \cdots \int_{-\pi}^\pi d \theta_n \, \prod_{l=1}^n
g(\theta_l) \prod_{1 \le j < k \le n} |e^{i \theta_k} -
e^{i \theta_j} |^\beta.
\end{equation}
As  the first step, guided by both the log-gas viewpoint and the
structure of (\ref{FH}), we conjecture the factorization
\begin{equation}\label{ff1}
{ Z_n^{(\beta)}[a(\theta) \prod_{j=1}^R |e^{i \theta} -
e^{i \phi_j} |^{q_j \beta} ] \over
Z_{n + \sum_{j=1}^R q_j}^{(\beta)}[1]  } 
\sim e^{- \sum_{j=1}^R q_j \log a(\theta_j)}
{ Z_n^{(\beta)}[a(\theta)]
\over
Z_{n + \sum_{j=1}^R q_j}^{(\beta)}[1]  }
{ Z_n^{(\beta)}[\prod_{j=1}^R |e^{i \theta} -
e^{i \phi_j} |^{q_j \beta} ] \over
Z_{n + \sum_{j=1}^R q_j}^{(\beta)}[1]  }
\end{equation}

From the work of Johansson \cite{Jo88,Jo98}, with the Fourier expansion
of $\log a(\theta)$ specified by (\ref{1.6a}) and assuming the coefficients
satisfy (\ref{3}), it has been proved for general $\beta > 0$ that
\begin{equation}\label{ff8}
{Z_{n+Q}^{(\beta)}[a(\theta)] \over Z_{n+Q}^{(\beta)}[1] } \sim
e^{c_0(n+Q)} e^{(2/\beta) \sum_{k=1}^\infty k c_k c_{-k} }.
\end{equation}
Regarding the second ratio on the right hand side of (\ref{ff1}), as first
noted in \cite{FP92,Fo92} and revised in \cite{FFGW03a}, the log-gas
viewpoint suggests that for $n \to \infty$ we have the factorization
\begin{equation}\label{ff2}
\prod_{1 \le j < k \le R} |e^{i \theta_k} - e^{i \theta_j}|^{\beta q_j q_k}
{ Z_n^{(\beta)}[\prod_{j=1}^R |e^{i \theta} -
e^{i \phi_j} |^{q_j \beta} ] \over
Z_{n + \sum_{j=1}^R q_j}^{(\beta)}[1]  }
\sim
\prod_{j=1}^R { Z_n^{(\beta)}[\prod_{j=1}^R |e^{i \theta} -
e^{i \phi_j} |^{q_j \beta} ] \over
Z_{n + q_j}^{(\beta)}[1]  }
\end{equation}
The large-$n$ expansion of a ratio closely related to the product on the
right hand side of (\ref{ff2}) is known for $\beta$ rational, in
particular
\begin{equation}\label{rr}
\beta/2 = s/r, \qquad s \: {\rm and} \: r \: {\rm relatively} \: {\rm prime}.
\end{equation}
Thus we have \cite{Fo92}
\begin{equation}\label{Aq0}
\prod_{j=1}^R { Z_n^{(\beta)}[\prod_{j=1}^R |e^{i \theta} -
e^{i \phi_j} |^{q_j \beta} ] \over
Z_{n}^{(\beta)}[1] } \sim n^{q^2 \beta/2} A_q,
\end{equation}
where
\begin{equation}\label{Aq}
A_q := r^{-q^2\beta/2} \prod_{\nu=0}^{r-1} \prod_{p=0}^{s-1}
{G^2(q/r+\nu/r-p/s+1) \over
G(2q/r+\nu/r-p/s+1) G(\nu/r-p/s+1) }.
\end{equation}
Finally, the formula for $C_{n,\beta}$ in (\ref{e.3}) together with
Stirling's formula shows
\begin{equation}
{Z_{n}^{(\beta)}[1] \over Z_{n+q}^{(\beta)}[1] } \sim
(\Gamma(\beta/2+1))^q (n\beta/2)^{-q\beta/2}.
\end{equation}

Combining the above results gives the sought $\beta$-generalization
of the Fisher-Hartwig formula in the case $b_r=0$.

\begin{conj}\label{cj1}
Let $\beta$ be rational and of the form (\ref{rr}), and let $a(\theta)$ be
as assumed for the validity of (\ref{ff8}). 
For $q_j \beta > - 1$ we expect
\begin{equation}\label{3.12}
\Big 
\langle \prod_{l=1}^N \Big (
a(\theta_l) \prod_{j=1}^R |e^{i\theta_l} - e^{i \phi_j} |^{q_j \beta}
\Big )
\Big \rangle_{{\rm C}\beta{\rm E}_n} 
\mathop{\sim}\limits_{n \to \infty}
e^{c_0(n + \sum_{j=1}^R q_j)} n^{(\beta/2) \sum_{j=1}^R q_j^2} E^{(\beta)}
\end{equation}
where, with $A_q$ specified by (\ref{Aq}),
\begin{equation}
E^{(\beta)} = e^{-\sum_{j=1}^R q_j \log a(\theta_j)}
e^{(2/\beta)\sum_{k=1}^\infty k c_k c_{-k}}
\prod_{1 \le j < k \le R} |e^{i \theta_k} - e^{i \theta_j}|^{-\beta q_j q_k}
\prod_{j=1}^R A_{q_j}.
\end{equation}
\end{conj}

It is of interest to extend Conjecture \ref{cj1} to include a factor
\begin{equation}\label{3.14}
\prod_{j=1}^R e^{-i(\beta/2)b_r {\rm arg} \, e^{i(\phi_j+\pi - \theta)}}
\end{equation}
in the average, and so obtain a $\beta$-generalization of the 
Fisher-Hartwig formula for general parameters. Although we don't have a
log-gas interpretation of the factor (\ref{3.14}), the case $R=1$
substituted in (\ref{3.12}) with $a(\theta)=$ gives an average which can
be evaluated in closed form, and the corresponding asymptotics computed
for rational $\beta$. This together with the structure of the original
Fisher-Hartwig formula (\ref{FH}), (\ref{FH1}) allows us to formulate the
sought $\beta$-generalization.

Now, by rotational invariance, independent of the value of $\phi$
\begin{eqnarray}\label{3.15}
\Big \langle \prod_{l=1}^n
e^{- i (\beta/2) b {\rm arg} \, e^{i (\phi + \pi - \theta_l)}}
| e^{i \theta_l} - e^{i \phi} |^{\beta q} \Big \rangle_{{\rm C}\beta {\rm E}_n}
& = & 
\Big \langle \prod_{l=1}^n
e^{ i \beta b \theta_l /2} | 1 + e^{i \theta_l} |^{\beta q} \nonumber \\
\Big \rangle_{{\rm C}\beta {\rm E}_n} & = &
{Z_n^{(\beta)}[e^{ i \beta b \theta /2} | 1 + e^{i \theta} |^{\beta q}] \over
Z_n^{(\beta)}[1] }.
\end{eqnarray}
But, from the theory of the Selberg integral (see e.g.~\cite{Fo02}), we
know the right hand side of (\ref{3.15}) has the explicit gamma function
evaluation
\begin{equation}\label{3.16}
{f_n(2cq,c) \over f_n(c(q+b),c) f_n(c(q-b),c)} \qquad {\rm where} \quad 
f_n(\alpha,c) := \prod_{j=0}^{n-1} {(\alpha + jc)! \over (jc)!}, \quad
c:= \beta/2.
\end{equation}
For $c \in \zz^+$ it was shown in \cite{Fo92} that
\begin{equation}\label{3.17}
f_n(\alpha,c) 
\mathop{\sim}\limits_{n \to \infty} \exp(\alpha n \log n) c^{\alpha n}
e^{- \alpha n} n^{-(c-1)\alpha/2 + \alpha^2/2c}
\prod_{p=0}^{c-1} {G(-p/c+1) \over G((\alpha-p)/c+1) },
\end{equation}
while for $r$ and $s$ relatively prime
\begin{equation}\label{3.18}
f_{rn}(\alpha,s/r) = \prod_{\nu=0}^{r-1}
{f_n(\alpha + s\nu/r,s) \over f_n(s\nu/r,s) }.
\end{equation}
Using (\ref{3.18}) and (\ref{3.17}) in (\ref{3.16}) it follows that for
$\beta$ rational of the form (\ref{rr}),
\begin{equation}\label{qb}
\Big \langle \prod_{l=1}^n e^{ i \beta b \theta_l/2} 
| 1 + e^{i \theta_l} |^{\beta q}
\Big \rangle_{{\rm C}\beta {\rm E}_{rn}}
\mathop{\sim}\limits_{n \to \infty} (rn)^{(\beta/2) (q^2 - b^2)} A_{q,b}
\end{equation}
where
\begin{equation}\label{qb1}
A_{q,b} := r^{-(q^2-b^2)\beta/2}
\prod_{\nu=0}^{r-1} \prod_{p=0}^{s-1}
{G((q+b)/r+\nu/r-p/s+1) G^2((q-b)/r+\nu/r-p/s+1) \over
G(2q/r+\nu/r-p/s+1) G(\nu/r-p/s+1) }.
\end{equation}
Note that in the case $b=0$ this reduces to (\ref{Aq}), (\ref{Aq0}) as
it must.

Knowing how, from the Fisher-Hartwig formula (\ref{FH}), (\ref{FH1}),
to generalize from the case $R=1$, general parameters, and the case
general $R$ but $b_r=0$ $(r=1,\dots,R)$, to the case of general parameters
and general $R$ lets us use (\ref{3.12}) and (\ref{qb}) to formulate a
$\beta$-generalization of the Fisher-Hartwig formula for general
parameters.

\begin{conj}\label{cj1a}
Let $\beta$ be rational and of the form (\ref{rr}), and let $a(\theta)$ be
as assumed for the validity of (\ref{ff8}). We expect, for some range of
parameters $\{b_j\}$ and $\{q_j\}$, 
\begin{equation}\label{3.12a}
\Big
\langle \prod_{l=1}^n a(\theta_l) \prod_{j=1}^R 
e^{-i(\beta/2)b_j {\rm arg} \, e^{i(\phi_j+\pi - \theta_l)}}
|e^{i\theta_l} - e^{i \phi_j} |^{q_j \beta}
\Big \rangle_{{\rm C}\beta{\rm E}_n}
\mathop{\sim}\limits_{n \to \infty}
e^{c_0n } n^{(\beta/2) \sum_{j=1}^R (q_j^2 - b_j^2)} \tilde{E}^{(\beta)}
\end{equation}
where, with $A_{q,b}$ specified by (\ref{qb1}),
\begin{eqnarray}
\tilde{E}^{(\beta)} & = & e^{(2/\beta)\sum_{k=1}^\infty k c_k c_{-k}}
\prod_{r=1}^R e^{-(q_r+b_r) \log a_-(\theta_r) }
e^{-(q_r-b_r) \log a_+(\theta_r) } \nonumber \\
&& \times
\prod_{1 \le r \ne s \le R}
(1 - e^{i (\theta_s - \theta_r)})^{-\beta (q_r+b_r)(q_s-b_s)/2} 
\prod_{j=1}^R A_{q_j,b_j}.
\end{eqnarray}
\end{conj}

\section{Fisher-Hartwig asymptotics for averages over the orthogonal and
symplectic groups}
\setcounter{equation}{0}
A problem in mathematical physics which, along with the Ising correlations,
motivated the Fisher-Hartwig formula (\ref{FH}) is the impenetrable Bose gas
on a circle. If the circle has circumference length $L$, it was shown by
Lenard \cite{Le64} that the ground state density matrix
$\rho_{N+1}^C(x)$ has the Toeplitz determinant form
\begin{eqnarray}\label{D1}
\rho_{N+1}^C(x;0) & = & {1 \over L} \det [ a_{j-k}^C(x)
]_{j,k=1,\dots,N} \nonumber \\
a_l^C(x) & := & {1 \over 2 \pi} \int_{-\pi}^\pi |e^{2 \pi i x/L} +
e^{i \theta} | |1 + e^{i \theta} | e^{- i l \theta} \, d \theta.
\end{eqnarray}
The symbol in (\ref{D1}) is of the form (\ref{5}) with
\begin{equation}\label{D1'}
a(\theta) = 1, \quad R=2, \quad a_1 = a_2 = {1 \over 2}, \quad
b_1 = b_2 = 0, \quad \theta_1 = 0, \quad \theta_2 = 2\pi x/L.
\end{equation}
Now a fundamental issue relating to the Bose gas is the occupation
$\lambda_0$ of the zero momentum state, which quantifies the phenomenum
of Bose-Einstein condensation (see e.g.~\cite{FFGW03a}). 
In the present system, which is
translationally invariant, $\lambda_0$ is related to the density matrix
by the simple formula
\begin{equation}\label{D1a}
\lambda_0 = \int_0^L \rho_{N+1}^C(x;0) \, dx = L
\int_0^1  \rho_{N+1}^C(LX;0) \, dX.
\end{equation}
For fixed $0< X < 1$ one thus seeks the $L \to \infty$ asymptotic form of
$\rho_{N+1}^C(LX;0)$. In an unpublished work as of 1968, made available to
the authors of \cite{FH68} and subsequently published in 1972 \cite{Le72},
Lenard obtained for the sought expansion
\begin{equation}\label{D1b}
 \rho_{N+1}^C(LX;0) \sim \rho_0 {G^4(3/2) \over \sqrt{2 \pi} }
\Big ( {\pi \over N \sin (\pi X) } \Big )^{1/2}
\end{equation}
where $\rho_0$ denotes the bulk density. 
Lenard obtained (\ref{D1b}) as an upper bound, which soon after was shown 
to be attained by Widom \cite{Wi73}.
Applying the Fisher-Hartwig
formula (\ref{FH}) with variables (\ref{D1'}) reproduces (\ref{D1b}).

According to (\ref{e.1}) the Toeplitz formula (\ref{D1}) can equivalently
be written as the $U(N)$ average
\begin{equation}\label{D2}
\rho_{N+1}^C(x;0) = {1 \over L} \Big \langle \prod_{l=1}^N
\Big |2 \sin \Big ( {\pi x \over L} - {\theta_l \over 2} \Big ) \Big |
\, 
\Big |2 \sin {\theta_l \over 2} \Big | \Big \rangle_{U(N)}.
\end{equation}
The study of the ground state density matrices for the impenetrable Bose
gas on a line of length $L$ with Dirichlet or Neumann boundary
conditions leads to formulas analogous to (\ref{D2}), only now the
averages are with respect to the eigenvalue probability density
functions for the classical groups $Sp(N)$ and $O^+(2N)$ respectively
(see e.g.~\cite{Fo02} for the specification of these PDFs). Thus one
has \cite{FFGW03a}
\begin{eqnarray}\label{3.6}
\rho_{N+1}^D(x;y) & = & {2 \over L} 
\sin {\pi x \over L} \sin {\pi y \over L}
\Big \langle \prod_{l=1}^N
\Big |2 (\cos  {\pi x \over L} - \cos \theta_l) \Big |
\Big |2 (\cos  {\pi y \over L} - \cos \theta_l) \Big | 
\Big \rangle_{Sp(N)} \nonumber  \\
\rho_{N+1}^N(x;y) & = & {1 \over 2L} 
\Big \langle \prod_{l=1}^N 
\Big |2 (\cos  {\pi x \over L} - \cos \theta_l) \Big |
\Big |2 (\cos  {\pi y \over L} - \cos \theta_l) \Big |
\Big \rangle_{O^+(2N)}.
\end{eqnarray}
Impenetrable bosons on the interval $[0,L]$ with Dirichlet boundary
conditions at $x=0$ and Neumann boundary conditions at $x=L$ also relate
to  a classical group. Thus from the fact that the single particle wave
functions are given by
$$
\phi_k^M(x) = \sqrt{2 \over L} \sin {\pi (k-1/2) x \over L}, \qquad
(k=1,2,\dots)
$$
(the superscript $M$ stands for ``mixed'') we see that the ground state
wave function
$$
\psi_0^M(x_1,\dots,x_N) = {1 \over \sqrt{N!}} \Big | \det
[ \phi_k^M(x_j) ]_{j,k=1,\dots,N} \Big |
$$
has the product form
$$
\psi_0^M(x_1,\dots,x_N) = {1 \over \sqrt{N!}}
\Big ( {1 \over 2 \sqrt{L}} \Big )^N \prod_{l=1}^N 2 \sin(\pi x_l/2L)
\prod_{1 \le j < k \le N} 2 | \cos \pi x_k/L - \cos \pi x_j/L |.
$$
The square of this quantity coincides with the eigenvalue PDF of the
classical group $O^+(2N+1)$ with $\theta = \pi x/L$ (for this we
ignore the fixed eigenvalue at $\theta = 0$). From this fact, as in
the derivation of (\ref{3.6}) detailed in \cite{FFGW03a}, it follows that
\begin{equation}\label{3.6a}
\rho_{N+1}^M(x;y)  =  {2 \over L}
\sin {\pi x \over 2L} \sin {\pi y \over 2L}
\Big \langle \prod_{l=1}^N
\Big |2 (\cos  {\pi x \over L} - \cos \theta_l) \Big |
\Big |2 (\cos  {\pi y \over L} - \cos \theta_l) \Big |
\Big \rangle_{O^+(2N+1)}.
\end{equation}

Using a combination of analytic calculations based on the Selberg
correlation integral \cite{Ka93}, and physical arguments based on
log-gas analogies, the large $N$, fixed $x/L,\, y/L, \, N/L$ limit of the
density matrices (\ref{3.6}) was computed in \cite{FFG03} to be equal to
\begin{equation}\label{3.6b}
\rho_{N+1}^D(x;y) \sim \rho_{N+1}^N(x;y) \sim \rho
{G^4(3/2) \over \sqrt{2N} }
{(X(1-X))^{1/8}(Y(1-Y))^{1/8} \over |X - Y|^{1/2} }
\Big |_{X = (1 + \cos \pi x/L)/2 \atop Y = (1 + \cos \pi y/L)/2}.
\end{equation}
Here we will show how recent rigorous asymptotic analysis \cite{BE02}
of Toeplitz $+$ Hankel determinants (\ref{1.9a}) with Fisher-Hartwig
type symbols can be used to prove that $\rho_{N+1}^M(x;y)$
exhibits the same asymptotic form (\ref{3.6b}). We will also show
how a result of \cite{BE02} can be used to confirm the asymptotic
form of a more general class of averages over $O^+(2N+1)$ which
can be deduced from a conjecture in \cite{FFG03}, and how this
conjecture in turn can be used to predict analogous asymptotics in
the case of averages over $Sp(N)$ and $O^+(2N)$. 

To begin we require a simple to verify identity noted in \cite{BR01}.

\begin{lemma}\label{lem1}
Suppose $g(\theta) = g(-\theta)$ and set
$g_j = {1 \over 2 \pi} \int_{-\pi}^\pi g(\theta) e^{-i j \theta} \, d
\theta$. We have
\begin{equation}\label{3.6c}
\det [g_{j-k} + g_{j+k+1} ]_{j,k=0,\dots, N-1} = 
\Big \langle \prod_{j=1}^N g(\theta_j) \Big \rangle_{O^-(2N+1)}
= \Big \langle \prod_{j=1}^N g(\pi - \theta_j)) \Big \rangle_{O^+(2N+1)}.
\end{equation}
\end{lemma}
Note that by the assumption on $g(\theta)$ the matrix in
(\ref{3.6c}) is symmetric. Also, the average in (\ref{3.6a}) is an even
function of $\theta_1$ and corresponds to the special case
\begin{eqnarray}\label{4.6}
g(\theta) & = & 2 | \cos {\pi x \over L} - \cos \theta | \,
2 | \cos {\pi y \over L} - \cos \theta | \nonumber \\
 & = &
\Big ( |2 - 2\cos (\theta - {\pi x \over L}) |
|2 - 2\cos (\theta + {\pi x \over L}) | 
|2 - 2\cos (\theta - {\pi y \over L}) |
|2 - 2\cos (\theta + {\pi y \over L}) | \Big )^{1/2}
\end{eqnarray} 
of (\ref{3.6c}). We observe that (\ref{4.6}) is an example of a symbol of
the form (\ref{5}). Fortunately, recent rigorous works \cite{BE01,BE02}
have determined the asymptotic form of the Hankel $+$ Toeplitz determinant
in (\ref{3.6c}) for all symbols (\ref{5}), with the restriction that for
$g(\theta)$ even (the case of interest in relation to (\ref{3.6c})),
$\theta_r \ne 0, \, \pm \pi$. Let us recall the result of
\cite[Thm.~6.1]{BE02}, simplified so that it relates to the even case of
(\ref{5}) with each $b_r=0$.

\begin{theorem}\label{th1}
Let
\begin{equation}\label{3.11'}
\log g(\theta) = \log a(\theta) + \sum_{r=1}^R a_r \Big (
\log (2|\cos\theta - \cos \theta_r| )
\Big ),
\end{equation}
where $a(\theta)$ is an even periodic function with the property that the
Fourier expansion of its logarithm (\ref{1.6a}) satisfies (\ref{3}),
together with some technical assumptions (for the latter, which may
not be necessary, see \cite{BE02}). We have
\begin{equation}\label{3.12b}
\det [g_{j-k} + g_{j+k+1} ]_{j,k=0,\dots, N-1} \sim
e^{(N + \sum_{j=1}^R a_j) c_0 }
(2N)^{\sum_{r=1}^R a_r^2} E
\end{equation}
where
\begin{eqnarray}\label{E}
E & = & \prod_{r=1}^R {G^2(1+a_r) \over G(1+2 a_r)}
e^{{1 \over 2} \sum_{k=1}^\infty k c_k^2 + \sum_{k=1}^\infty c_{2k-1} }
e^{-\sum_{j=1}^N a_j \log a(\theta_j) } \nonumber \\
&& \times \prod_{r=1}^R {|1 - e^{i \theta_r}|^{a_r} \over
|1 + e^{i \theta_r}|^{a_r} |1 - e^{2 i \theta_r}|^{a_r^2} }
\prod_{1 \le r < s \le R} \Big (
|1 - e^{i(\theta_r - \theta_s)}|
|1 - e^{i(\theta_r + \theta_s)}| \Big )^{-2 a_r a_s}.
\end{eqnarray}
\end{theorem}

Recalling (\ref{3.6c}) it follows from Theorem \ref{th1} that
\begin{eqnarray}
&&
\Big \langle \prod_{j=1}^N |2(\cos {\pi x \over L} - \cos \theta_j)| \,
|2(\cos {\pi y \over L} - \cos \theta_j)| 
\Big \rangle_{O^-(2N+1)} \nonumber \\
&& \sim (2N)^{1/2} G^4(3/2)
{|1+e^{\pi i x/L}|^{1/4} \over |1-e^{\pi i x/L}|^{3/4} }
{|1+e^{\pi i y/L}|^{1/4} \over |1-e^{\pi i y/L}|^{3/4} }
{1 \over |1 - e^{\pi i (x-y)/L}|^{1/2} |1 - e^{\pi i (x+y)/L}|^{1/2} }
\end{eqnarray}
Substituting this in (\ref{3.6a}) shows
\begin{eqnarray}
\rho_{N+1}^M(x,y) & \sim & {(2N)^{1/2} \over 2L} G^4(3/2)
{|1-e^{2 \pi i x/L}|^{1/4} |1-e^{2 \pi i y/L}|^{1/4} \over
|1 - e^{\pi i (x-y)/L} |^{1/2} |1 - e^{\pi i (x+y)/L} |^{1/2} }
\nonumber \\
& = &
\rho
{G^4(3/2) \over \sqrt{2N} }
{(X(1-X))^{1/8}(Y(1-Y))^{1/8} \over |X - Y|^{1/2} }
\Big |_{X = (1 + \cos \pi x/L)/2 \atop Y = (1 + \cos \pi y/L)/2},
\end{eqnarray}
thus rigorously establishing the asymptotic form (\ref{3.6b}) derived, but
not rigorously proved, in \cite{FFG03}
 for the cases of Dirichlet and Neumann boundary
conditions.

The eigenvalue distributions for $Sp(N)$, $O^+(2N)$, $O^-(2N+1)$ and
$O^+(2N+1)$ are proportional to
\begin{equation}\label{un.1}
\prod_{l=1}^N(1+\cos \theta_l)^{\lambda_1} (1 - \cos \theta_l)^{\lambda_2}
\prod_{1 \le j < k \le N} (\cos \theta_k - \cos \theta_j)^2, \qquad
0 \le \theta_l \le \pi
\end{equation}
for $(\lambda_1,\lambda_2) = (1,1), \, (0,0), \, (1,0)$ and $(0,1)$
respectively (our convention is not to include the delta function
corresponding to a fixed eigenvalue, nor the delta functions corresponding
to the conjugate eigenvalues). 
Thus to obtain the asymptotics of the averages in (\ref{3.6})
it is sufficient to obtain the asymptotics of
\begin{equation}\label{un.2}
\Big \langle \prod_{l=1}^N \prod_{r=1}^R \Big (2 | \cos \theta_l -
\cos \phi_r | \Big )^{2 a_r} \Big \rangle_{C_N(\lambda_1,\lambda_2)}
\end{equation}
where $C_N(\lambda_1,\lambda_2)$ refers to the normalized form of
(\ref{un.1}). In the cases $(\lambda_1, \lambda_2) = (1,0)$ or 
$(0,1)$, due to the identity (\ref{3.6c}), we can read off the
asymptotic form from (\ref{3.12b}). But for general $(\lambda_1,\lambda_2)$
the asymptotic form of (\ref{un.2}) is not included in Theorem \ref{th1}.
Instead, we will use a conjecture from \cite{FFG03} to formulate the
result.

Let us first recall the conjectured asymptotic form from \cite{FFG03}.
Define
\begin{equation}\label{od}
H_{n,\lambda_1,\lambda_2}[f(x)] :=
\int_0^1 dx_1 \cdots \int_0^1 dx_n \, \prod_{l=1}^n f(x_l) x_l^{\lambda_1}
(1 - x_l)^{\lambda_2} \prod_{1 \le j < k \le n}
|x_k - x_j|^2.
\end{equation}
Then the argument given in \cite{FFG03} predicts\footnote{Unfortunately there
are a number of inaccuracies in the reporting of the conjecture in
\cite{FFG03}. The term   $H_{n+\sum_{j=1}^Rq_j,\lambda_1,\lambda_2}[1]$
in the denominator on the left hand side of (\ref{od1}) has mistakenly
been written as $H_{n,\lambda_1,\lambda_2}[1]$ in equations
(90), (94), (96) and (97); the factors $\prod_{j=1}^Ry_j^{-\lambda_1 q_j}
(1 - y_j)^{-\lambda_2 q_j}$ are missing and should be paired with
$\prod_{1 \le j < k \le R} |y_k - y_j|^{-2 q_j q_k}$ throughout;
and the term $e^{-(\lambda_1 + \lambda_2)[h(0)+h(1)]/4}$ in (96)
and (97) should read
$e^{-(\lambda_1 h(0) + \lambda_2 h(1))/2}$.}
\begin{eqnarray}\label{od1}
&&{H_{n,\lambda_1,\lambda_2}[e^{h(x)} \prod_{r=1}^R |y_r - x|^{2 q_r}]
\over H_{n+\sum_{j=1}^Rq_j,\lambda_1,\lambda_2}[1]} \nonumber \\
&&\qquad \sim
\exp \Big [ {n + \sum_{r=1}^R q_r + (\lambda_1 + \lambda_2)/2 \over \pi}
\int_0^1 {h(x) \over [x(1-x)]^{1/2} } \, dx \Big ]
\exp \Big [ \sum_{r=1}^R (-q_r + q_r^2) \log 2n \Big ] K \nonumber \\
\end{eqnarray}
where 
\begin{eqnarray}
K & = & \prod_{j=1}^Ry_j^{-\lambda_1 q_j}
(1 - y_j)^{-\lambda_2 q_j} \prod_{1 \le j < k \le R} |y_k - y_j|^{-2 q_j q_k}
e^{-(\lambda_1 h(0) + \lambda_2 h(1))/2} e^{-\sum_{r=1}^R q_r h(y_r)}
\nonumber \\
&& \times \exp \Big [ {1 \over 4 \pi^2} \int_0^1 dx \,
{h(x) \over (x(1-x))^{1/2} } \int_0^1 dy \,
{h'(y) (y(1-y))^{1/2} \over x - y} \Big ] \nonumber \\
&& \times \prod_{r=1}^R( y_r (1 - y_r))^{-{q_r}^2/2}
\prod_{r=1}^R {1 \over \pi^{q_r}}
{G^2(q_r+1) \over G(2q_r + 1) }.
\end{eqnarray}

In preparation for relating this to the average (\ref{un.2}) let
\begin{equation}\label{od3}
h\Big ({1 \over 2}(1 + \cos \theta) \Big ) = c_0 + 2 \sum_{n=1}^\infty c_n
\cos n \theta.
\end{equation}
We then have that
\begin{equation}\label{od4}
{1 \over \pi} \int_0^1 {h(x) \over (x(1-x))^{1/2} } \, dx =
{1 \over \pi} \int_0^\pi h\Big ({1 \over 2}(1 + \cos \theta)
\Big ) \, d \theta =
c_0
\end{equation}
\begin{equation}\label{od5}
{1 \over 2} (\lambda_1 h(0) + \lambda_2 h(1) ) =
{1 \over 2} (\lambda_1 + \lambda_2) c_0 + \sum_{n=1}^\infty c_n
(\lambda_1 + (-1)^n \lambda_2)
\end{equation}
while (\ref{od3}) together with the cosine expansion 
$$
\log (2 | \cos \theta - \cos \phi |) = - \sum_{n=1}^\infty
{2 \over n} \cos n \theta \cos n \phi
$$
shows
\begin{equation}\label{od6}
{1 \over 4 \pi^2} \int_0^1 dx \,
{h(x) \over (x(1-x))^{1/2}} \int_0^1 dy \,
{h'(y) (y(1-y))^{1/2} \over x - y}  = {1 \over 2}
\sum_{n=1}^\infty n c_n^2.
\end{equation}
Also, as noted in \cite{FFG03},
\begin{equation}\label{odp}
H_{n,a,b}[1] = {G(n+1+a) G(n+1+b) G(n+1+a+b) \over
G(1+a) G(1+b) G(2n+1+a+b) } G(n+2).
\end{equation}
Since \cite{Ba00}
\begin{equation}\label{Ba}
\log  {G(n+1+a) \over G(n+1+b)}
\mathop{\sim}\limits_{n \to \infty}
(b-a)n + {a-b \over 2} \log (2 \pi) + \Big ( (a-b)n
+ {a^2 - b^2 \over 2} \Big ) \log n + o(1)
\end{equation}
we deduce
\begin{equation}\label{od7}
{H_{n,a,b}[1] \over H_{n+Q,a,b}[1] } \sim
{2^{4nQ + 2Q^2 + 2Q(a+b)} \over (2 \pi n)^Q }.
\end{equation}
Finally we note that under the change of variables
$$
x_l = {1 \over 2} (1 + \cos \theta_l)
$$
the integrand in (\ref{od}) contains as a factor the
(unnormalized) eigenvalue probability density function (\ref{un.1}).
Explicitly, with $\tilde{h}(\theta) := h({1 \over 2}(1 + \cos \theta))$
we have
\begin{equation}\label{od8}
{H_{n,\lambda_1,\lambda_2}[e^{h(x)} \prod_{r=1}^R |y_r - x|^{2 q_r}]
\over H_{n,\lambda_1,\lambda_2}[1]} \Big |_{y_r = {1 \over 2}(1 +
\cos \phi_r) } =
\Big \langle \prod_{l=1}^n e^{\tilde{h}(\theta_l)}
\prod_{r=1}^R \Big | {1 \over 2} (\cos \phi_r - \cos \theta_l)
\Big |^{2 q_r} \Big \rangle_{C_N(\lambda_1, \lambda_2)}.
\end{equation}

Making use of (\ref{od4})--(\ref{od8}) shows that the asymptotic
formula (\ref{od1}) for the integral (\ref{od}) is equivalent to
an asymptotic formula generalizing Theorem \ref{th1}.

\begin{conj}\label{cj3}
Let $\log a(\theta)$ have the Fourier expansion (\ref{1.6a}), with
coefficients satisfying (\ref{3}). We expect that for $N \to \infty$
\begin{equation}\label{od9}
\Big \langle \prod_{l=1}^N a(\theta_l) \prod_{r=1}^R|2(\cos \phi_r - \cos
\theta_l) |^{2a_r} \Big \rangle_{C_N(\lambda_1 + 1/2, \lambda_2 + 1/2)}
\: \sim \: e^{(N + \sum_{r=1}^R a_r ) c_0 }
(2N)^{\sum_{r=1}^R a_r^2} \tilde{K}
\end{equation}
where, with $E$ specified by (\ref{E}),
\begin{equation}\label{od10}
\tilde{K} = \prod_{r=1}^R {1 \over |1 + e^{i \phi_r} |^{2(\lambda_1 - 1)a_r}
|1 + e^{i \phi_r} |^{2 \lambda_2 a_r} }
e^{-\sum_{n=1}^\infty c_n(\lambda_1 - 1 + (-1)^n \lambda_2) } E.
\end{equation}
\end{conj}

We can apply some checks to (\ref{od1}). As already remarked,
with $(\lambda_1,\lambda_2)=(1,0)$ the probability density function
$C_N(\lambda_1,\lambda_2)$ coincides with the eigenvalue
probability density function for $O^-(2N+1)$, and (\ref{od9}) must
coincide with (\ref{3.12b}), as indeed it does. Also, changing variables
$\theta_l \mapsto \pi - \theta_l$ and interchanging $\lambda_1$
and $\lambda_2$ leaves (\ref{un.1}) invariant, and thus the average
(\ref{un.2}) invariant if we also put $\phi_r \mapsto \pi - \phi_r$,
$a(\theta) \mapsto a(\pi - \theta)$ (and thus $c_n \mapsto (-1)^n c_n$).
Recalling the definition (\ref{E}) of $E$ we see that (\ref{od10})
exhibits this symmetry. Another check follows from a factorization
identity, relating an average over the unitary group to a product of
averages over the orthogonal and symplectic groups \cite{Wi92,BR01,Fo03}.

\begin{prop}\label{p.1}
With $g(\theta) = g(-\theta)$ we have
\begin{equation}\label{r1}
\Big \langle \prod_{l=1}^{2N+1} g(\theta_l) \Big \rangle_{U(2N+1)} =
\Big \langle \prod_{l=1}^{N+1} g(\theta_l) \Big \rangle_{O^+(2N+2)}
\Big \langle \prod_{l=1}^{N} g(\theta_l) \Big \rangle_{Sp(N)}
\end{equation}
\end{prop}

With
$$
g(\theta) = a(\theta) \prod_{r=1}^R (2 | \cos \theta - \cos \phi_r |)^{2a_r}
$$
in (\ref{r1}) we see that the conjectured asymptotic form (\ref{od9})
for the right hand side is consistent with the Fisher-Hartwig
formula (\ref{FH}). 

The identity (\ref{r1}) is also of interest for providing an exact
formula for the product of the density matrix in Dirichlet boundary
conditions and in Neumann boundary conditions. Thus recalling (\ref{3.6})
we see from (\ref{r1}) that
$$
{1 \over L^2} \sin {\pi x \over L} \sin {\pi y \over L} 
\Big \langle \prod_{l=1}^{2N+1} \Big | 2(\cos {\pi x \over L} -
\cos \theta_l) \Big | \Big |
2(\cos {\pi x \over L} -
\cos \theta_l) \Big | \Big \rangle_{U(2N+1)} =
\rho_{N+2}^N(x,y) \rho_{N+1}^D(x,y).
$$

\section{Impenetrable bosons in a harmonic trap and random matrix
averages over the GUE and LUE}
\setcounter{equation}{0}
From a physical viewpoint the most relevant setting for the impenetrable
Bose gas is confinement by a harmonic potential (see \cite{FFGW03a,
FFGW03b} and references therein). Then the ground state wave function
$\psi_0^H$ is proportional to
$$
\prod_{l=1}^N e^{-x_l^2/2} \prod_{1 \le j < k \le N} |x_k - x_j|,
$$
and $|\psi_0^H|^2$ is identical to the eigenvalue probability density
function for the Gaussian unitary ensemble of complex Hermitian matrices.
The combination of log-gas arguments and analytic calculation based on
the Selberg correlation integral used to analyze (\ref{od}) was used in
\cite{FFGW03b} to analyze the asymptotic form of
\begin{equation}\label{4.1}
e^{-\sum_{j=1}^R 2 N q_r y_r^2}
{G_{N,\sqrt{2N}}[ \prod_{r=1}^R |x - y_r|^{2 q_r} ] \over
G_{N + \sum_{r=1}^R q_r, \sqrt{2N} }[1] }
\end{equation}
where
\begin{equation}\label{z.a}
G_{N,a}[f(x)] 
:= \int_{-\infty}^\infty dx_1 \cdots \int_{-\infty}^\infty dx_N \,
\prod_{l=1}^N f(x_l) e^{-a^2 x_l^2} \prod_{1 \le j < k \le N} |x_k - x_j|^2,
\end{equation}
in the special case $R=2$, $q_1=q_2=1/2$ which specifies the ground state
density matrix. As our first point of interest we will generalize this
calculation to general $R$ and $q_r$ ($q_r > -1/2$).

The log-gas perspective \cite{Fo92} suggests the factorization
\begin{equation}\label{z.0}
\prod_{1\le j < k \le R} |y_j - y_k|^{2 q_j q_k}
e^{-\sum_{r=1}^R 2 N q_r y_r^2} 
{G_{N,\sqrt{2N} }[ \prod_{r=1}^R |x - y_r|^{2 q_r} ] \over
G_{N + \sum_{r=1}^R q_r, \sqrt{2N}}[1] }
\mathop{\sim}\limits_{N \to \infty}
\prod_{r=1}^R e^{- 2 N q_r y_r^2}
{G_{N,\sqrt{2N}}[  |x - y_r|^{2 q_r} ] \over
G_{N +  q_r,\sqrt{2N}}[1] }.
\end{equation}
Next, from the theory of Selberg correlation integrals \cite{Ka93},
for $q_r \in \zz_{\ge 0}$ we have the duality formula \cite{BF97} 
\begin{equation}\label{z.1}
{G_{N,a}[(x-y_r)^{2q_r}] \over G_{N,a}[1] } =
{G_{2q_r,a}[(y_r + i x)^N] \over G_{2q_r,a}[1] }.
\end{equation}
It is a fairly straight forward exercise, detailed in 
\cite[section 5.3]{BF97} for a related problem, to use the saddle point
method to compute the large $N$ expansion of the integral on the right hand
side of (\ref{z.1}). One finds
\begin{equation}\label{z.2}
e^{-2 N q_r x_r^2} G_{2q_r, \sqrt{2N} }[(y_r + i x)^N]
\mathop{\sim}\limits_{N \to \infty}
\Big ( {2q_r \atop q_r} \Big ) (G_{q_r,1}[1])^2 e^{-q_r N}
2^{-2q_rN} (4N)^{-q_r^2} (1 - y_r^2)^{q_r^2/2}.
\end{equation}
Also, we know (see e.g.~\cite{Fo02}) the exact evaluation 
\begin{equation}\label{4.35a}
G_{n,1}[1] = 2^{-n(n-1)/2} \pi^{n/2} G(n+2),
\end{equation}
which together with the asymptotic expansion (\ref{Ba}) implies
\begin{equation}\label{z.3}
{G_{N,\sqrt{2N}}[1] \over G_{N+q,\sqrt{2N}}[1] }
\mathop{\sim}\limits_{N \to \infty}
{2^{2N q + q^2 - q} \over \pi^q } e^{q N} N^{-q}.
\end{equation}
Combining (\ref{z.1})--(\ref{z.3}) gives the asymptotic formula
\begin{equation}\label{z.4}
e^{-2N q_r  y_r^2}
{G_{N,\sqrt{2N}}[|x - y_r|^{2q_r}] \over
G_{N+q_r, \sqrt{2N}}[1] }
\mathop{\sim}\limits_{N \to \infty}
{G^2(q_r+1) \over G(2q_r+1)} {2^{2q_r^2 - q_r} \over \pi^{q_r} }
N^{q_r^2 - q_r} (1 - y_r^2)^{q_r^2/2} 
\end{equation}
which we have proved for $q_r \in \zz_{\ge 0}$, and conjecture as
being valid for all $q_r > -1/2$. This same result can also be
deduced from results in \cite{BH00}. Substituting this in (\ref{z.0})
gives the sought asymptotic form of (\ref{4.1}).

\begin{conj}\label{ju1}
Let $q_r > -1/2$, and let $G_{N,a}[f]$ be given by (\ref{z.a}).
We expect
\begin{eqnarray}\label{k1}
&& 
e^{-\sum_{r=1}^R 2 N q_r y_r^2}
{G_{N,\sqrt{2N} }[ \prod_{r=1}^R |x - y_r|^{2 q_r} ] \over
G_{N + \sum_{r=1}^R q_r, \sqrt{2N}}[1] } \nonumber \\
&&
\qquad \mathop{\sim}\limits_{N \to \infty}
\prod_{1\le j < k \le R} |y_j - y_k|^{-2 q_j q_k}
\prod_{r=1}^R 
{G^2(q_r+1) \over G(2q_r+1)} {2^{2q_r^2 - q_r} \over \pi^{q_r} }
N^{q_r^2 - q_r} (1 - y_r^2)^{q_r^2/2}.
\end{eqnarray}
\end{conj}

An extension of (\ref{k1}) can also be formulated. Let $a(x)$ be analytic
on $[-1,1]$. Then it has rigorously been proved that \cite{Jo98}
\begin{eqnarray}\label{c1}
\lefteqn{
{G_{N,\sqrt{2N}}[e^{a(x)}] \over G_{N,\sqrt{2N}}[1] }} \nonumber
\\ &&
\mathop{\sim}\limits_{N \to \infty} \exp \Big ( {2N \over \pi}
\int_{-1}^1 a(x) \sqrt{1-x^2} \, dx \Big )
\exp \Big ( {1 \over 4 \pi^2}
\int_{-1}^1 dx \, {a(x) \over (1 - x^2)^{1/2} }
\int_{-1}^1 dy \, {a'(y) (1 - y^2)^{1/2} \over x - y} \Big ).
\end{eqnarray}
The structure of (\ref{od1}) in the case $\lambda_1 = \lambda_2 = 0$
suggests how (\ref{c1}) can be combined with (\ref{k1}) to generalize
the latter.

\begin{conj}\label{ju2}
Let $a(x)$ be analytic on $[-1,1]$. It is expected that
\begin{eqnarray}\label{c2}
&&
e^{-\sum_{r=1}^R 2 N q_r y_r^2}
{G_{N,\sqrt{2N} }[ e^{a(x)} \prod_{r=1}^R |x - y_r|^{2 q_r} ] \over
G_{N + \sum_{r=1}^R q_r, \sqrt{2N}}[1] }  \nonumber \\
&& \qquad
\mathop{\sim}\limits_{N \to \infty}
\Big ( {\rm RHS} \, (\ref{c1}) \Big )  \Big |_{N \mapsto N + \sum_{r=1}^R
q_r}
\Big ( {\rm RHS} \, (\ref{k1}) \Big )
e^{-\sum_{r=1}^R q_r a(y_r) }.
\end{eqnarray}
\end{conj}

We remark that in the special case $a(x) = kx$, Conjecture \ref{ju2} can
be reduced to Conjecture \ref{ju1}. To see this, use completion of squares
to note
$$
G_{N,\sqrt{2N}}\Big [e^{kx} \prod_{r=1}^R \Big |x - y_r \Big |^{2q_r}\Big ] =
e^{k^2/8} G_{N,\sqrt{2N}}\Big [\prod_{r=1}^R\Big 
|x + {k \over 2N} - y_r \Big |^{2q_r} \Big ].
$$
According to Conjecture \ref{ju1} we have
$$
e^{-\sum_{r=1}^R 2N q_r (y_r - k/4N)^2}
{G_{N,\sqrt{2N}}[e^{kx} \prod_{r=1}^R |x + {k \over 2N} - y_r|^{2q_r}] \over
G_{N + \sum_{r=1}^R q_r, \sqrt{2N}}[1] }
\mathop{\sim}\limits_{N \to \infty}
{\rm RHS} \, (\ref{c1})
$$
and thus
$$
e^{-\sum_{r=1}^R 2N q_r y_r^2} 
{G_{N,\sqrt{2N}}[e^{kx} \prod_{r=1}^R |x - y_r|^{2q_r}] \over
G_{N + \sum_{r=1}^R q_r, \sqrt{2N}}[1] }
\sim e^{k^2/8} \, {\rm RHS} \, (\ref{c1}) \, e^{- \sum_{r=1}^R q_r y_r}.
$$
Since (\ref{od6}) gives
$$
{1 \over 4 \pi^2} \int_{-1}^1 dx \, {a(x) \over (1-x^2)^{1/2}}
\int_{-1}^1 dy \, {a'(y) (1 - y^2)^{1/2} \over x - y}
\Big |_{a(x) = kx} = {k^2 \over 8}
$$
this is in agreement with (\ref{c2}).

Let us now turn our attention to a variation of the impenetrable Bose gas
in a harmonic well, which also has the features of being related to a
random matrix ensemble. In reduced units, the Hamiltonian for the system
is
\begin{equation}\label{H1}
H = - \sum_{j=1}^N {\partial^2 \over \partial x_j^2} +
\sum_{j=1}^N \Big ( a'(a'-1) {1 \over x_j^2} + x_j^2 \Big ), \qquad x_j>0.
\end{equation}
Thus in addition to the harmonic well,
the particles are restricted to the half line by a repulsive potential 
(requiring $a'>1$)
at
the origin proportional to $1/r^2$. This is the non-interacting case of the
so called type $B$ Calogero-Sutherland Hamiltonian \cite{Ya94}, 
for which the interacting case has $1/r^2$ pair repulsion.
The ground
state wave function for (\ref{H1}) is proportional to
\begin{equation}\label{4.12'}
\prod_{l=1}^N e^{-x_l^2/2} (x_l^2)^{a'/2}
\prod_{1 \le j < k \le N} |x_k^2 - x_j^2|.
\end{equation}
We recognize the square of the ground state wave function as being
identical to the probability density function for the singular values
of $n \times N$ complex Gaussian matrices with $a'=n-N+1/2$
(see e.g.~\cite[Ch.~2]{Fo02}. 
Changing variables $x_l^2 \mapsto x_l$ this is refered to as
the Laguerre unitary
ensemble.
The problem of computing the asymptotic
form of the density matrix for this system suggests analyzing the
asymptotic form of the more general quantity
\begin{equation}\label{L0}
\prod_{r=1}^R y_r^{2a'q_r} e^{-4N q_r y_r^2}
{L_N[\prod_{r=1}^R |x^2 - y_r^2|^{2q_r}] \over
L_{N+\sum_{r=1}^R q_r}[1] }
\end{equation}
where
\begin{equation}\label{L1}
L_N[f] := \int_0^\infty dx_1 \cdots \int_0^\infty dx_N \,
\prod_{l=1}^N f(x_l) x_l^{2a'} e^{-4N x_l^2}
\prod_{1 \le j < k \le N} |x_k^2 - x_j^2|^2,
\end{equation}
in analogy with (\ref{4.1}).

From a log-gas perspective, the integrand in (\ref{L1}) corresponds to a
one-component system interacting on the half line $x>0$, subject to a
one-body confining potential $2N x^2 - (a' - 1/2) \log x$.  In addition
to the electrostatic energy $-\log |x-x'|$ at the point $x$ due to the
interaction with a charge at $x'$, there is also a term $-\log|x+x'|$
due to the interaction with an image charge at $-x'$ (outside the system,
since $x'>0$). In keeping with the image charge interpretation, for
each charge at $x$ one 
requires a term $-{1 \over 2} \log |2x|$ to account for the interaction
between a charge and its own image (the factor of $1/2$ is because this
energy is shared between the charge and its image, the latter being
outside the system). From this viewpoint we can interpret
$$
\prod_{1\le j < k \le R} |y_k^2 - y_j^2|^{2 q_j q_k}
\prod_{r=1}^R (2y_r)^{q_r^2} e^{-q_r 4 N y_r^2}
|y_r|^{(2a'-1)q_r}
{L_N[\prod_{r=1}^R |x^2 - y_r^2|^{2q_r}] \over
2^{\sum_{r=1}^R q_r} L_{N+\sum_{r=1}^R q_r}[1] }
$$
as a ratio of partition functions for log-gas systems, and analogous to
(\ref{z.0}) we expect the factorization into
\begin{equation}\label{L6}
\prod_{r=1}^R  (2y_j)^{q_j^2}  e^{-q_r 4 N y_r^2}
|y_r|^{(2a'-1)q_r} 
{L_N[|x^2 - y_r^2|^{2q_r}] \over
2^{q_r} L_{N+ q_r}[1] } 
\end{equation}
for $N \to \infty$. 

To analyze (\ref{L6}) in the limit $N \to \infty$ we
make the change of variables $x_l^2 \mapsto x_l$ and introduce
\begin{equation}\label{5.16'}
\tilde{L}_{N,c}[f] :=
\int_0^\infty dx_1 \cdots \int_0^\infty dx_N \,
\prod_{l=1}^N f(x_l) x_l^{a'-1/2} e^{-c x_l}
\prod_{1 \le j < k \le N} (x_k - x_j)^2
\end{equation}
so that it reads
\begin{equation}\label{5.16a}
\prod_{r=1}^R  (2y_r)^{q_r^2}  e^{-q_r 4 N y_r^2}
y_r^{(2a'-1)q_r}
{\tilde{L}_{N,4N}[|x - y_r^2|^{2q_r}] \over
\tilde{L}_{N+ q_r,4N}[1] }.
\end{equation}
To proceed further, we use the fact that for $q \in \zz_{\ge 0}$ we have the
duality formula \cite{Fo94}
\begin{eqnarray}\label{L7}
\lefteqn{
{\tilde{L}_{N,c}[|x-t|^{2q}] \over \tilde{L}_{N,c}[1] \Big |_{a \mapsto
a + 2q} }  = {1 \over M_{2q}(a,N) } } \nonumber \\
&& \times 
\int_{-1/2}^{1/2} dx_1 \cdots \int_{-1/2}^{1/2} dx_{2q} \,
\prod_{l=1}^{2q} e^{\pi i x_l (a-N) }
|1 + e^{2 \pi i x_l} |^{a+N} e^{-ct e^{2 \pi i x_l}}
\prod_{1 \le j < k \le 2q} | e^{2 \pi i x_k} - e^{2 \pi i x_j} |^2 \nonumber \\
\end{eqnarray}
where on the right hand side $a=a'-1/2$ and
\begin{eqnarray}\label{L8}
M_n(a,b) & : = & \int_{-1/2}^{1/2} dx_1 \cdots \int_{-1/2}^{1/2} dx_n \,
\prod_{l=1}^n e^{\pi i x_l(a-b) } |1 + e^{2 \pi i x_l} |^{a+b}
\prod_{1 \le j < k \le n} |e^{2 \pi i x_k} - e^{2 \pi i x_j} |^2
\nonumber \\
& = & {G(n+1+a+b) \over G(1+a+b)} {G(1+a) \over G(n+1+a)}
{G(1+b) \over G(n+1+b)} G(n+2)
\end{eqnarray}
(for the last equality see e.g.~\cite{Fo02}).
If we suppose temporarily that $a \in Z_{\ge 0}$, the right hand side of
(\ref{L7}) with $c=4N$ can be written as the contour integral
\begin{equation}\label{4.16a}
{1 \over M_{2q}(a,-N) } \int_{\cal C} {dz_1 \over 2 \pi i z_1}
\cdots \int_{\cal C} {dz_{2q} \over 2 \pi i z_{2q}} \,
\prod_{l=1}^{2q} (1 + z_l)^a (1 + 1/z_l)^N e^{-4N tz_l}
\prod_{1 \le j < k \le 2q} (z_k - z_j) (1/z_k - 1/z_j)
\end{equation}
where ${\cal C}$ is any simple closed contour which encircles the origin.
To analyze this for $N \to \infty$, following \cite{Fo94}  where the case
$q=1$ was considered, we note the $N$-dependent terms in the integrand
have a stationary point when
\begin{equation}\label{L8a}
z = z_{\pm} := - {1 \over 2} \pm i {1 \over 2}(1/t - 1)^{1/2}.
\end{equation}
By deforming the contour ${\cal C}$ to pass through $z_+$ for $q$ of
the integrations, and to pass through $z_-$ for the remaining $q$
integrations, we readily deduce from the representation (\ref{4.16a})
of (\ref{L7}) that
\begin{eqnarray}\label{L9}
\lefteqn{e^{-q_r 4N y_r^2} y_r^{(2a'-1)q_r} (2y_r)^{q_r^2}
\tilde{L}_{N,4N}[|x - y_r^2|^{2q_r}]} \nonumber \\
&& = {\tilde{L}_{N,4N}[1] |_{a' \mapsto a' +2 q_r} \over
M_{2q_r}(a'-1/2,N) } e^{-q_r 4N y_r^2} y_r^{(2a'-1)q_r} (2y_r)^{q_r^2}
\nonumber \\
&& \qquad \times
\Big ( {2q_r \atop q_r} \Big ) e^{-4N  y_r^2 q_r (z_+ + z_-) +
N q_r \log |1 + 1/z_+|^2} {1 \over |z_+|^{4q_r^2} } |z_+ - z_-|^{2q_r^2}
|1 + z_+|^{q_r(2a'-1)} \Big ( {1 \over 2 \pi} \Big )^{2 q_r} \nonumber \\
&&  \qquad \times
\Big | {N \over 2} \Big ( {1 \over z_+^2} - {1 \over (1 + z_+)^2}
\Big ) \Big |^{-q_r^2} ( G_{q_r}[1])^2.
\end{eqnarray}
Now, with $t=y_r^2$ in (\ref{L8a}) 
\begin{eqnarray}
z_+ + z_- = -1, \qquad |1 + {1 \over z_+} |^2 = 1, \qquad
|z_+ - z_-|^2 = \Big ( {1 \over y_r^2} - 1 \Big ) \nonumber \\
|1 + z_+|^2 = |z_+|^2 = {1 \over 4 y_r^2}, \qquad
\Big | {1 \over z_+^2} - {1 \over (1 + z_+)^2} \Big | =
16 y_r^4 \Big ( {1 \over y_r^2} - 1 \Big )^{1/2}
\end{eqnarray}
so the right hand side of (\ref{L9}) simplifies to
\begin{equation}\label{L9c}
{\tilde{L}_{N,4N}[1] |_{a' \mapsto a' +2 q_r} \over
M_{2q_r}(a'-1/2,N) } N^{-q_r^2} \Big ( {1 \over 2 \pi} \Big )^{2q_r}
\Big ( {2q_r \atop q_r} \Big ) 2^{2q_r^2}
2^{-q_r(2a'-1)} (G_{q_r}[1])^2
(1 - y_r^2)^{q_r^2/2}.
\end{equation}
Furthermore we know (see e.g.~\cite{Fo02})
\begin{equation}
\tilde{L}_{N,c}[1] = c^{-N^2 - N(a'-1/2)} {G(N+2) G(a'+N +1/2) \over
G(a'+1/2) },
\end{equation}
and making use too of (\ref{L8a}) it follows from the asymptotic expansion
(\ref{Ba}) that
\begin{equation}\label{4.35b}
{\tilde{L}_{N,4N}[1] |_{a' \mapsto a' +2 q_r} \over \tilde{L}_{N+q_r,4N}[1]
M_{2q_r}(a'-1/2,N) } N^{-q_r^2} \sim
2^{2(q_r^2 + q_r (a'-1/2) )} N^{q_r^2 - q_r}.
\end{equation}
Substituting (\ref{4.35b}) in (\ref{L9c}), evaluating $G_{q_r}[1]$ therein
according to (\ref{4.35a}) and simplifying we obtain the
$N \to \infty$ expansion
\begin{equation}
e^{-q_r 4N y_r^2} y_r^{(2a'-1)q_r} (2y_r)^{q_r^2}
{ {L}_{N}[|x^2 - y_r^2|^{2q_r}] \over 2^{q_r} L_{N+q_r}[1] }
\sim {G^2(q_r+1) \over G(2q_r+1) } {2^{3 q_r^2 - q_r} \over \pi^{q_r} }
N^{q_r^2 - q_r} (1 - y_r^2)^{q_r^2/2},
\end{equation}
proved for $q_r \in \zz_{\ge 0}$ and expected to be true for all
$q_r > -1/2$. Substituting this in (\ref{L6}) gives, as a conjecture,
the sought asymptotic form of (\ref{L0}). 

\begin{conj}\label{c6}
For $N \to \infty$, and assuming $q_r > - 1/2$ for each $r=1,\dots,R$,
\begin{eqnarray}\label{4.25}
\lefteqn{
\prod_{r=1}^R y_r^{2a'q_r} e^{-4N q_r y_r^2}
{L_N[\prod_{r=1}^R |x^2 - y_r^2|^{2 q_r}] \over L_{N+\sum_{r=1}^R q_r}[1]}
} \nonumber \\
&& \sim
\prod_{1 \le j < k \le R} |y_k^2 - y_j^2|^{-2 q_j q_k}
\prod_{r=1}^R {G^2(q_r+1) \over G(2q_r+1)}
{2^{2q_r^2} \over \pi^{q_r}} N^{q_r^2 - q_r} y_r^{-q_r^2 + q_r}
(1 - y_r^2)^{q_r^2/2}.
\end{eqnarray}
\end{conj}

It is of interest to extend (\ref{4.25}) in an analogous way to how
(\ref{c2}) extends (\ref{k1}).  First we use (\ref{od1}) with $q_r=0$,
and (\ref{c1}) to conjecture that for $a(x)$ analytic on $[0,1]$
\begin{eqnarray}\label{4.26}
\lefteqn{
{L_N[e^{a(x)}] \over L_N[1]} \sim
\exp \Big ( {4 (N + (2a'-1)/4) \over \pi} \int_0^1 a(x) \sqrt{1 - x^2}
\, dx \Big ) }
 \nonumber \\
&& \times 
 \exp \Big ( {1 \over \pi^2} \int_0^1 dx \,
{a(x) \over (1 - x^2)^{1/2}} \int_0^1 dy \, {y a'(y) (1 - y^2)^{1/2} \over
x^2 - y^2} \Big ) e^{-a' a(0)/2}.
\end{eqnarray}
Combining this with (\ref{4.25}) as in (\ref{c2}) gives us the LUE
analogue of Conjecture \ref{ju2}.

\begin{conj}\label{ju3}
Let $a(x)$ be analytic on $[0,1]$. It is expected that
\begin{eqnarray}\label{4.27}
\lefteqn{
\prod_{r=1}^R y_r^{2a' q_r} e^{-4N q_r y_r^2}
{L_N[e^{a(x)} \prod_{r=1}^R|x^2 - y_r^2|^{2q_r} ] \over
L_{N + \sum_{r=1}^R q_r}[1] } } \nonumber \\
&& \mathop{\sim}\limits_{N \to \infty}
\Big ( {\rm RHS} \, (\ref{4.26}) \Big ) \Big |_{N \to N + \sum_{r=1}^R q_r}
\Big ( {\rm RHS} \, (\ref{4.25}) \Big ) e^{-\sum_{r=1}^R q_r a(y_r)}.
\end{eqnarray}
\end{conj}

We can check the consistency of (\ref{c2}) and (\ref{4.27}). For this we
make use of a factorization identity analogous to Proposition \ref{p.1}
\cite{Fo03}

\begin{prop}
Let $g(\theta) = g(-\theta)$. We have
\begin{equation}\label{4.28}
{G_{2N,a}[g(x)] \over G_{2N,a}[1] } =
{L_{N,a}^{(0)}[g(x)] \over L_{N,a}^{(0)}[1]}
{L_{N,a}^{(2)}[g(x)] \over L_{N,a}^{(2)}[1]}
\end{equation}
where
$$
L_{N,a}^{(p)}[g(x)] := \int_{-\infty}^\infty dx_1 \cdots
\int_{-\infty}^\infty dx_N \, \prod_{l=1}^N g(x_l) |x_l|^p e^{-a^2 x_l^2}
\prod_{1 \le j < k \le N} (x_k^2 - x_j^2)^2.
$$
\end{prop}

Let $a(x)$ be even and choose
$$
a= \sqrt{4N}, \qquad g(x) = e^{a(x)} 
\prod_{r=1}^R (x^2 - y_r^2)^{2q_r}. 
$$
According to Conjecture \ref{ju2},
\begin{eqnarray}
&& e^{-2a^2\sum_{r=1}^R q_r y_r^2} {G_{2N,a}[g(x)] \over G_{2(N + \sum_{r=1}^R
q_r),a}[1] }  \nonumber 
\\
&&\quad \mathop{\sim}\limits_{N \to \infty}
\exp \Big ( {4 \over \pi} (2N + 2 \sum_{r=1}^R q_r)
\int_0^1 a(x) \sqrt{1 - x^2} \, dx \Big ) \nonumber \\&& \quad \times
\exp \Big ( {1 \over \pi^2}
\int_0^1 dx \,
{a(x) \over (1 - x^2)^{1/2}} \int_0^1 dy \, {y a'(y) (1 - y^2)^{1/2} \over
x^2 - y^2} \Big ) \nonumber \\
&&
\qquad \times \prod_{1 \le j < k \le R} |y_j^2 - y_k^2|^{-4 q_j q_k}
\prod_{r=1}^R  (1 - y_r^2)^{q_r^2} |2y_r|^{-2q_r^2}
\nonumber \\
&&
\qquad \times \Big ( \prod_{r=1}^R {G^2(q_r+1) \over G(2q_r+1)}
{2^{2q_r^2 - q_r} \over \pi^{q_r} } (2N)^{q_r^2 - q_r} \Big )^2
e^{-2 \sum_{r=1}^R q_r a(y_r) }.
\end{eqnarray}

For the right hand side of (\ref{4.28}) as implied by Conjecture
\ref{ju3} to be consistent with this we require
\begin{equation}\label{vf}
2^{4 \sum_{r=1}^R q_r}
{G_{2N,\sqrt{4N}}[1] \over L_N[1] |_{a'=0}  L_N[1] |_{a'=1} } \sim
{G_{2(N+\sum_{r=1}^R q_r),\sqrt{4N}}[1] \over 
L_{N +\sum_{r=1}^R q_r}[1] |_{a'=0}  L_{N+\sum_{r=1}^R q_r}[1] |_{a'=1}} .
\end{equation}
But the method of derivation of (\ref{4.28}) given in \cite{Fo03} shows
that for general $n$, 
$$
{G_{2n,\sqrt{4n}}[1] \over
L_{n} |_{a'=0} L_{n} |_{a'=1} } = 2^{2n} {(2n)! \over (n!)^2} \sim
{2^{4n} \over (\pi n)^{1/2} },
$$
verifying (\ref{vf}).

Let us now apply Conjecture \ref{ju3} to the calculation of the density
matrix $\rho_{N+1}^L(x,y)$ for the state (\ref{4.12'}) with $N+1$
particles,
\begin{eqnarray}
\lefteqn{\rho_{N+1}^L(x,y)  := {N+1 \over C_{N+1} } e^{-x^2/2 - y^2/2} 
(xy)^{a'}
} \nonumber \\
&& \times 
\int_0^\infty dx_1 \cdots \int_0^\infty dx_N \,
\prod_{l=1}^N x_l^{2a'} e^{-x_l^2} |x^2 - x_l^2| |y^2 - y_l^2|
\prod_{1 \le j < k \le N} (x_k^2 - x_j^2)^2
\end{eqnarray}
where
$$
C_{N+1} := \int_0^\infty dx_1 \cdots  \int_0^\infty dx_{N+1}
\prod_{l=1}^{N+1} x_l^{2a'} e^{-x_l^2} 
\prod_{1 \le j < k \le N+1} (x_k^2 - x_j^2)^2.
$$
In terms of the average (\ref{L1}) we thus have
$$
2 \sqrt{N} \rho_{N+1}^L(2 \sqrt{N} X,2 \sqrt{N} Y) = (N+1)
e^{-2N X^2 - 2NY^2} (XY)^{a'}
{L_N[\prod_{l=1}^N |x^2 - X^2| |x^2 - Y^2|] \over L_{N+1}[1] }.
$$
On the right hand side we can apply Conjecture \ref{ju3} with $R=2$,
$q_1=q_2=1/2$ and so obtain the asymptotic form
\begin{equation}\label{laf}
2 \sqrt{N} \rho_{N+1}^L(2 \sqrt{N} X,2 \sqrt{N} Y)  \sim
2 \sqrt{N} {G^4(3/2) \over \pi}
{(XY)^{1/4} \over |X^2 - Y^2|^{1/2} } (1 - X^2)^{1/8} (1 - Y^2)^{1/8}.
\end{equation}

The asymptotic form (\ref{laf}) can in turn be used to specify the
occupations $\lambda_j$ of the  low-lying effective single particle states
$\phi_j$, which by definition satisfy the eigenvalue equation
\begin{equation}\label{se}
\int \rho_N(x,y) \phi_j(y) \, dy = \lambda_j \phi_j(x).
\end{equation}
Thus, with $x=2\sqrt{N} X$, $y=2\sqrt{N} Y$ and $j$ fixed, introducing the
scaled effective single particle states \cite{Pa03,FFGW03a}
$$
(4N)^{1/2} \phi_j(x) \mapsto \varphi_j(X),
$$
substituting (\ref{se1}) and using the fact that $ \rho_N^L(x,y)$
is supported on $x,y \in [0,2\sqrt{N}]$ we obtain the explicit
integral equation
\begin{equation}\label{se1}
2 \int_0^1 {X^{1/4} (1 - X^2)^{1/8} \varphi_j(X) \over
|X^2 - Y^2|^{1/2} } \, dX = \bar{\lambda}_j
{ \varphi_j(Y) \over Y^{1/4} (1 - Y^2)^{1/8} }
\end{equation}
where
\begin{equation}\label{se2}
\lambda_j =  \sqrt{N} {G^4(3/2) \over \pi} \bar{\lambda}_j.
\end{equation}
We see immediately that the occupations of the low-lying effective
single particle states are proportional to $\sqrt{N}$, as has been
found for the impenetrable Bose gas in periodic boundary conditions
\cite{Le64,FFGW03a}, in a harmonic trap \cite{Pa03,FFGW03b} and in
Dirichlet and Neumann boundary conditions \cite{FFG03}.
An appropriate analysis similar to that undertaken in 
\cite[Appendix B]{FFGW03b} gives the same upper bound on
$\bar{\lambda}_0$ as 
found for the same quantity in the
case of the harmonic trap \cite{FFGW03b}, but a detailed analysis of
(\ref{se1}) remains.

\section{Concluding remarks}
\setcounter{equation}{0}
\subsection{Universal form for Hankel asymptotics}
Analogous to (\ref{e.1}), Hankel determinants are related to log-gas
partition functions according to the formula
\begin{eqnarray}\label{6.0}
\lefteqn{
\det[a_{j+k}]_{j,k=0,\dots,n-1} } \nonumber \\
&& \quad = {1 \over n!} \int_{-\infty}^\infty dx_1 \, e^{-nV(x_1)}
\cdots \int_{-\infty}^\infty dx_n \, e^{-nV(x_n)} 
\prod_{l=1}^n a(x_l) \prod_{1 \le j < k \le n} (x_k - x_j)^2
\nonumber \\
&& \quad =: A_n(e^{-nV(x)})[a(x)]
\end{eqnarray}
where
$$
a_p = \int_{-\infty}^\infty a(x) x^p e^{-nV(x)} \, dx.
$$
For $V(x)$ an even degree polynomial independent of $n$ with positive
leading coefficient and no real zeros, it was proved by Johansson
\cite{Jo98} that
\begin{eqnarray}\label{6.1}
\lefteqn{
{A_n(e^{-nV(x)})[e^{a(x)}] \over A_n(e^{-nV(x)})[1]} }\nonumber \\
&& \mathop{\sim}\limits_{n \to \infty}
\exp \Big ( n \int_{c_1}^{c_2} a(x) \rho(x) \, dx \Big )
\exp \Big ( {1 \over 4 \pi^2} \int_{c_1}^{c_2} dx
{a(x) \over \sqrt{(x-c_1)(c_2-x)} }
 \int_{c_1}^{c_2} dy  {a'(y)  \sqrt{(y-c_1)(c_2-y)} \over x-y}
\Big ). \nonumber \\
\end{eqnarray}
Here $\rho(x)$ is the scaled density in the log-gas system corresponding
to $A_n(e^{-nV(x)})[1]$, supported on $[c_1,c_2]$ and normalized so that
$$
\int_{c_1}^{c_2} \rho(x) \, dx = 1.
$$
The asymptotic formula (\ref{c1}) corresponds to the special case 
$V(x) = {1 \over 2} x^2$, $\rho(x) = {2 \over \pi} \sqrt{1 - x^2}$ of
(\ref{6.1}). To extend Conjecture \ref{ju2} to more general $V$ this
suggests we simply write the latter in terms of $\rho(x)$. 

\begin{conj}\label{c9}
Under the conditions of the validity of (\ref{6.1}) we expect
\begin{eqnarray}\label{x1}
&&e^{-n \sum_{r=1}^R q_r V(y_r)}
{A_n(e^{-nV(x)})[e^{a(x)}\prod_{j=1}^r |x - y_j|^{q_j}] 
\over A_{n+ \sum_{j=1}^R q_j} (e^{-nV(x)})[e^{a(x)}] } \nonumber \\
&&
\qquad \mathop{\sim}\limits_{n \to \infty}
e^{-\sum_{r=1}^R q_r a(y_r)}
\prod_{1\le j < k \le R} |y_k - y_j|^{-2q_j q_k}
\prod_{r=1}^R {G^2(q_r+1) \over G(2q_r+1) }
(2 \pi N)^{q_r^2 - q_r} (\rho(y_r))^{q_r^2}.
\end{eqnarray}
\end{conj}

We remark that in the case $R=1$, $e^{a(x)}=1$, this conjecture (together with
some corroborative analysis) was formulated earlier by Br\'ezin and Hikami
\cite{BH00} (see also \cite{SF02}).

Conjecture \ref{ju3} can similarly be extended, although we work with the
quantity (\ref{5.16'}) in favour of (\ref{L1}) so as to have a Hankel
determinant interpretation according to (\ref{6.0}). In the log-gas system
corresponding to (\ref{5.16'}) one has $\rho(x) = {2 \over \pi x^{1/2}}(1-
x)^{1/2}$. Recalling the equality between (\ref{L6}) and (\ref{5.16a}),
and writing $y_r^2 \mapsto y_r$, $a(x^{1/2}) \mapsto a(x)$ we see that
Conjecture \ref{ju3} can be rewritten to imply 
\begin{eqnarray}\label{4.27a}
\lefteqn{
\prod_{r=1}^R y_r^{(a'-1/2) q_r} e^{-4N q_r y_r}
{\tilde{L}_{N,4N}[e^{a(x)} \prod_{r=1}^R|x - y_r|^{2q_r} ] \over
\tilde{L}_{N + \sum_{r=1}^R q_r,4N}[e^{a(x)}] } } \nonumber \\
&& \mathop{\sim}\limits_{N \to \infty}
e^{-\sum_{r=1}^R q_r a(y_r)}
\prod_{1\le j < k \le R} |y_k - y_j|^{-2q_j q_k}
\prod_{r=1}^R {G^2(q_r+1) \over G(2q_r+1) }
(2 \pi N)^{q_r^2 - q_r} (\rho(y_r))^{q_r^2},
\end{eqnarray}
thus assuming the universal form (\ref{x1}) and suggesting the following
analogue of (\ref{6.1}) and Conjecture \ref{c9}.

\begin{conj}\label{cj8}
Let $V(x)$ be a polynomial independent of $n$, with positive leading 
coefficient and no real zeros on $[0,\infty)$. Let
\begin{equation}
\tilde{A}_n(x^\alpha e^{-nV(x)})[a(x)] :=
{1 \over n!} \int_0^\infty dx_1 \, x_1^\alpha e^{-nV(x_1)}
\cdots \int_0^\infty dx_n \, x_n^\alpha  e^{-nV(x_n)}
\prod_{l=1}^n a(x_l) \prod_{1 \le j < k \le n} (x_k - x_j)^2.
\end{equation}
Analogous to (\ref{6.1}) we expect that
\begin{eqnarray}\label{6.1a}
\lefteqn{
{\tilde{A}_n(x^\alpha e^{-nV(x)})[e^{a(x)}] \over 
\tilde{A}_n(x^\alpha e^{-nV(x)})[1]} } \nonumber \\
&& \mathop{\sim}\limits_{n \to \infty}
\exp \Big ( n \int_{0}^{c_2} a(x) \rho(x) \, dx \Big )
\exp \Big ( {1 \over 4 \pi^2} \int_{0}^{c_2} dx \,
{a(x) \over \sqrt{x(c_2-x)} }
 \int_{c_1}^{c_2} dy \, {a'(y)  \sqrt{y(c_2-y)} \over x-y}
\Big )
\end{eqnarray}
where $\rho(x)$ is the scaled density in the log-gas corresponding to
$\tilde{A}_n(x^\alpha e^{-nV(x)})[1]$, with support on $[0,c_2]$.
Furthermore, with the same meaning of $\rho(x)$, we expect
\begin{equation}
\prod_{r=1}^R y_r^\alpha e^{-n  q_r V(y_r)}
{\tilde{A}_n(x^\alpha
e^{-nV(x)})[e^{a(x)}\prod_{j=1}^r |x - y_j|^{q_j}]
\over \tilde{A}_{n+ \sum_{j=1}^R q_j}  
(x^\alpha e^{-nV(x)})[e^{a(x)}] } \mathop{\sim}\limits_{n \to \infty}
{\rm RHS} \, (\ref{x1}).
\end{equation}
\end{conj}

As a final comment on this point, we note that the universal form given
by the right hand side of (\ref{x1}) is also exhibited by the Fisher-Hartwig
formula (\ref{FH}). Thus, with $z_r := e^{i \theta_r}$ we see that
$$
{D_n[e^{a(\theta)} \prod_{r=1}^R |e^{i \theta} - z_r|] \over
D_{n+\sum_{j=1}^R}[e^{a(\theta)} ] }
\mathop{\sim}\limits_{n \to \infty}
{\rm RHS} \, (\ref{x1}) \Big |_{y_r = z_r \atop \rho(y) = N/2 \pi}.
$$

\subsection{Further Toeplitz $+$ Hankel structures}
The identity (\ref{3.6c}) of Lemma \ref{lem1} has counterparts for
averages over $Sp(N)$ and $O^+(2N)$ \cite{BR01}.

\begin{lemma}
Suppose $g(\theta) = g(-\theta)$, set $g_j = {1 \over 2 \pi}
\int_{-\pi}^\pi g(\theta) e^{-i j \theta} \, d \theta$, and let
$C_N(\lambda_1,\lambda_2)$ refer to the normalized form of (\ref{un.1}).
We have
\begin{eqnarray}\label{u5}
\det [a_{j-k} + a_{j+k}]_{j,k=0,\dots,N-1}
& = &
\Big \langle \prod_{j=1}^N g(\theta_j) \Big \rangle_{O^+(2N)} \: = \:
\Big \langle \prod_{j=1}^N g(\theta_j) \Big \rangle_{C_N(0,
0)} \nonumber \\
\det [a_{j-k} - a_{j+k+2}]_{j,k=0,\dots,N-1}
& = &
\Big \langle \prod_{j=1}^N g(\theta_j) \Big \rangle_{Sp(N)} \: = \:
\Big \langle \prod_{j=1}^N g(\theta_j) \Big \rangle_{C_N(1,
1)}.
\end{eqnarray}
\end{lemma}

Choosing $g(\theta)$ as in (\ref{3.11'}), Conjecture \ref{cj3} gives the
asymptotic behaviour of the right hand sides in (\ref{u5}), and thus the
conjectured form of these Toeplitz $+$ Hankel structures.

\subsection{Fluctuation formula perspective and future directions}
Let $p:=p(x_1,\dots,x_N)$ be an $N$-dimensional probability density
function. The stochastic quantity $A= \sum_{j=1}^N a(x_j)$, with the
$\{x_j\}$ sampled from $p$, is referred to as a linear statistic. Its
distribution $P_A(t)$ is defined by
\begin{equation}\label{gm1}
P_A(t) = \Big \langle \delta\Big (t - \sum_{j=1}^N a(x_j) \Big )
\Big \rangle_p,
\end{equation}
and taking the Fourier transform of this gives
\begin{equation}\label{gm2}
\tilde{P}_A(k) =  \Big \langle \prod_{j=1}^N e^{i k a(x_j)}
 \Big \rangle_p.
\end{equation}

The structure of the average (\ref{gm2}) is common to the averages studied
in this paper. As an illustration of the content of the asymptotic
formulas from this viewpoint, consider Johansson's result (\ref{ff8}).
Written in terms of the average (\ref{e.1}) with $g(\theta) = 
e^{i k a(\theta)}$, it reads
\begin{equation}\label{gm3}
D_n^{(\beta)}[e^{i k a(\theta)}]
\mathop{\sim}\limits_{n \to \infty} e^{i k c_0 n}
e^{-(2/\beta) k^2 \sum_{n=1}^\infty n c_n c_{-n} }
\end{equation}
where $\{c_n\}_{n=0,\pm1,\dots}$ are the Fourier coefficients in the
expansion of $a(\theta)$,
\begin{equation}\label{gm3a}
a(\theta)  = \sum_{n=-\infty}^\infty c_n e^{i n \theta}.
\end{equation}
A key feature of the exponents in the exponentials on the right hand side
of (\ref{gm3}) is that they form a quadratic polynomial in $k$. Thus
substituting this in (\ref{gm1}) and taking the inverse transform gives the
Gaussian distribution
\begin{equation}\label{gm3b}
P_A(t) \mathop{\sim}\limits_{n \to \infty} 
{1 \over (2 \pi \sigma^2)^{1/2} } e^{-(t - \mu)^2/2 \sigma^2}
\end{equation}
with
\begin{equation}\label{gm4}
\mu = n c_0, \qquad \sigma^2 = {4 \over \beta} \sum_{p=1}^\infty p
c_p c_{-p}.
\end{equation}
As noted by  Johansson \cite{Jo88}, in the case $\beta = 2$ this gives
a Gaussian fluctuation formula interpetation of Szeg\"o's theorem.
A perculiar feature is that although the mean is proportional to $n$,
the variance is $O(1)$, so fluctuations are strongly suppressed. It
is formulas of the type (\ref{gm3b}), (\ref{gm4}) which led to the
successful theoretical explanation of the phenomenom of universal
conductance fluctuations in mesoscopic wires (see e.g.~\cite{Be97}), in
which the conductance --- an order $N$ quantity --- is written as a linear
statistic of certain eigenvalues and is shown to have $O(1)$ fluctuations
with variance given by an analytic formula of the type (\ref{gm4}).

All our generalizations of the Fisher-Hartwig formula involve a term
of the form $e^{Q^2 \log n}$ as the first correction to the leading order
behaviour $e^{c_0n}$. However again when written as an average of the type
(\ref{gm2}) the exponential of a quadratic in $k$ again results.
Consider for example (\ref{3.12}). With $\{c_n\}$ specified by
(\ref{gm3a}) we have
$$
\Big \langle e^{i k a (\theta) + i k \beta \sum_{j=1}^R q_j
\log | e^{i \theta} - e^{i \phi_j} | } \Big \rangle_{{\rm C}\beta{\rm E}_n}
\mathop{\sim}\limits_{n \to \infty}
e^{i k c_0 n} e^{- k^2 (\beta/2) ( \sum_{j=1}^R q_j^2) \log n}
$$
and thus, as first noted in \cite{BF97a}, with
$$
A = \sum_{l=1}^N \Big ( a(\theta_l) + \beta \sum_{j=1}^R q_j
\log | e^{i \theta_l} - e^{i \phi_j} | \Big )
$$
the asymptotic form of the corresponding distribution is given by the
Gaussian (\ref{gm3b}) with
$$
\mu = n c_0, \qquad \sigma^2 = \beta \Big ( \sum_{j=1}^R q_j^2 \Big )
\log n.
$$
Thus the variance diverges logarithmically. This class of Gaussian
fluctuation theorem has found use in the application of random matrix
theory to the study of the statistical properties of the zeros of the
Riemann zeta function \cite{KS00,HKO00}. The study of the  statistical 
properties of the zeros of families of
$L$-functions requires averages over the different classical groups 
\cite{KS00a,C03,H03,KS03}. We might anticipate that our new results of
Section 4 will find application in this topic.

Of course it remains to prove the conjectures of this paper. Of
these, Conjecture \ref{cj1a} is the most general, as it involves
Fisher-Hartwig type parameters $\{q_j\}$, $\{b_j\}$ as well as the
log-gas type parameter $\beta$. It is also of interest to extend
Conjectures \ref{cj3}, \ref{c9} and  \ref{cj8} to this level of 
generality. Another direction of generality is to extend the
domain of integration from a circle or line to a two-dimensional
region \cite{Fo98,Ri03}.

\section*{Acknowledgement}
The financial support of the Australian Research Council is
acknowledged.

\end{document}